\documentclass[reprint,aps,pre,superscriptaddress,floatfix,longbibliography]{revtex4-1}

\usepackage{soul}

\usepackage{graphicx}
\usepackage{array}
\usepackage{color}
\usepackage{amsmath}
\usepackage{amsxtra}
\usepackage{amstext}
\usepackage{amssymb}
\usepackage{amsthm}
\usepackage{mathtools}
\usepackage{bm}
\usepackage{hyperref}
\usepackage{xcolor}
\usepackage{latexsym}
\usepackage{dsfont}

\newtheorem{theorem}{Theorem}
\newtheorem{prop}[theorem]{Proposition}

\newcommand{\conv}{{\rm conv}}
\newcommand{\Cr}{{C_{\rm r}}}
\newcommand{\dist}{{\rm dist}}
\newcommand{\Sr}{S_{\rm reg}}
\newcommand{\vol}{{\rm vol}}

\begin{document}

\renewcommand{\figurename}{Figure}

\title{On the Robustness of Democratic Electoral Processes to Computational Propaganda}

\author{Glory M. Givi}
\affiliation{Department of Quantum Matter Physics, University of Geneva, CH-1211 Geneva, Switzerland}
\affiliation{School of Engineering, University of Applied Sciences of Western Switzerland HES-SO, Sion CH-1951, Sion, Switzerland}
\author{Robin Delabays}
\email{Corresponding author.}
\affiliation{School of Engineering, University of Applied Sciences of Western Switzerland HES-SO, Sion CH-1951, Sion, Switzerland}
\author{Matthieu Jacquemet}
\affiliation{School of Engineering, University of Applied Sciences of Western Switzerland HES-SO, Sion CH-1951, Sion, Switzerland}
\affiliation{Department of Mathematics, University of Fribourg, CH-1700, Fribourg, Switzerland}
\author{Philippe Jacquod}
\affiliation{Department of Quantum Matter Physics, University of Geneva, CH-1211 Geneva, Switzerland}
\affiliation{School of Engineering, University of Applied Sciences of Western Switzerland HES-SO, Sion CH-1951, Sion, Switzerland}

\date{\today}

\begin{abstract}
There is growing evidence of systematic attempts to influence democratic elections by controlled and digitally organized dissemination of fake news.
This raises the question of the intrinsic robustness of democratic electoral processes against external influences. Particularly interesting is to identify the social characteristics 
of a voter population that renders it more resilient against opinion manipulation. Equally important is to determine which of the existing 
democratic electoral systems is
more robust to external influences. 
Here we construct a mathematical electoral model to address these two questions. 
We find that, not unexpectedly, biased electorates with clear-cut elections are overall quite resilient against opinion manipulations,
because inverting the election outcome requires to change the opinion of many voters. 
More interesting are 
unbiased or weakly biased electorates with close elections. We find that such populations
are more resilient against opinion manipulations
(i) if they are less polarized and (ii) when voters interact more with each other, regardless of their opinion differences, and that
(iii) electoral systems based on proportional representation
are generally the most robust. 
Our model qualitatively captures the volatility of the US House of Representatives elections. We 
take this as a solid validation of our approach. 
\end{abstract}

\maketitle

\section*{Introduction}

Social networks have become unavoidable tools in modern democratic elections.
They have dramatically changed the way politicians interact with their electorate,
allowing  campaigns to connect with electors, mobilize 
supporters and activists, and spread their message faster, more easily and over a larger range than ever before. From that point of view, social networks are new 
communication platforms that strengthen the democratic dialogue. Simultaneously, by their very structure and dynamic, social networks facilitate the fast and massive dissemination 
of information that is often hard, if not impossible to verify. As a matter of fact, there is a darker side to social networks and their use in democratic processes, which goes by the name of {\it computational propaganda}~\cite{sanovich2017a,woolley2017a,How2018a,How2018b}. This novel kind of political manipulation is based on bots --  software-created and -operated, automated social media accounts. Large 
number of bots generate colossal amounts of contents so fast, that it is virtually impossible to determine what is right and what is not. Indeed, fake news appear legitimate when repeated sufficiently often, 
because human perception is that so many "people" cannot be all lying simultaneously.
Bots manipulate online discussions and change people's perceptions of political issues and entities~\cite{Met2012}. There are accumulating 
examples of massive uses of bots in recent political campaigns. 
Bots seem to have been used intensively to try and influence people's perception of UK's role in the European Union during the 2016 Brexit campaign~\cite{How2016}.
In the 2016 US presidential election, social media platforms have been used to spread large amounts of false information~\cite{Bessi2016,How2018b}. 
In the 2017 UK general election, the Labour Party used bots to spread electoral messages on Tinder~\cite{How2018b}, while the incumbent candidate Jair Bolsonaro
massively used social media to spread targeted information in the 2018 Brazilian presidential election~\cite{arnaudo2017a}. Both the ruling Bharatiya Janata Party and the Indian National Congress widely used online misinformation in their campaign strategies in the 2019 Indian general election~\cite{Sye2022}. Last but not least, false claims of democrats stealing the election were disseminated by
the incumbent candidate Donald Trump and his supporters, following the 2020 US presidential election. There are several other known cases. 

Social media users make about 60 \% of the world's population,
therefore it can be expected that attempts to manipulate elections through social networks are going to multiply. While the true impact of computational propaganda on past
elections is hard to measure, evidence exist that emotions can be transferred through social media~\cite{Kramer2014} and in particular from bots to humans~\cite{Cai2023}.
One may therefore expect that well targeted computational propaganda efforts can have a significant effect on election results. As but one example, the Labour Party won several 
electoral constituencies by narrow margins in the 2017 UK general election. This has been attributed to computational propaganda efforts by some~\cite{How2018b}.
The main efforts against the emergence of computational propaganda and the associated threat to democratic processes worldwide
has been to try and construct fast and efficient ways to identify bots from standard human accounts, and to
regulate social networks into banning bot accounts.
Two altogether different lines of action in the fight against computational propaganda are (i) to try and identify the social characteristics of an electorate that makes it more resilient against opinion manipulation and (ii)
to determine which of the existing democratic electoral systems is the most robust to external influences. These are the two issues we investigate in this paper. 
To the best of our knowledge, these important questions have rarely been addressed within the framework of computational social science~\cite{Laz09,Con12}, with the very recent exception of Raducha et al.~\cite{Rad23}.

To that end, we construct a mathematical model of computational social science, where voters have an opinion encoded in a multi-dimensional vector agent. Each component of that vector
measures the affinity that the voter has for a given political party. In our dynamical model, these
components evolve according to the voter's natural opinion -- the one they would have if they were left in isolation -- and social interactions among voters. In the spirit of the Deffuant
~\cite{Deffuant2000} and Hegselmann-Krause~\cite{Hegselmann2002} models of opinion dynamics, these interactions exist only between pairs of voters with opinion vectors within a certain distance in 
opinion space. That distance is a parameter of the model, mimicking the openness of the electorate to differing opinions.
The mathematical model determines the voter final opinions.
The outcome of an election is next extracted from the latter through a procedure that reflects different electoral processes.

External influence from computational propaganda is then modeled as a targeted shift in some agents' natural opinion.
The magnitude of this shift is increased, until the electoral outcome changes. 
A qualitative measure of robustness of the election process is the magnitude of the effort needed to change the election outcome: more robust social populations and electoral systems do not change their vote until larger manipulation efforts are applied. We determine the dependence of that robustness for different electoral processes and populations with different social characteristics. The latter 
are encoded in the distribution of natural opinions -- allowing to tune between populations with homogeneous to polarized opinions, to multi-party populations, with or without electoral bias towards one or several parties -- and in the opinion distance over which voters interact with one another -- modeling the tolerance and openness they have to different views and opinions. 

Our model amplifies on but goes far beyond existing voter and opinion dynamics models, such as Abelson's~\cite{Abelson1967} or Taylor's~\cite{Taylor1968} models. 
It encompasses several characteristics that were present only individually in some models so far,
such as the finite interaction distance between agents~\cite{Deffuant2000,Hegselmann2002}, opinion vectors with dimension larger than two~\cite{Deffuant2000,Lorenz2008,Nedic2012,Friedkin2018,Stamoulas2018,Baumann2020a}, 
each of which representing a potential party affiliation, and the natural opinion that a voter would have, were they not interacting with  other voters~\cite{Taylor1968}.
This latter characteristic, in particular, protects our results from  
artificial features of opinion dynamics models with finite confidence bound, such as consensus and fragmentation~\cite{Chen2020}, that are unrealistic in the present context. It furthermore allows to model 
different populations with varying degrees of opinion polarization and initial partisan bias, and to investigate the influence of the latter on electoral processes and their robustness. One key new ingredient
in our model is that it generalizes consensus-like dynamics to multidimensional spaces of agents, where each dimension corresponds to one of several political parties.
Non-trivial is to order parties from left to right without biasing in favor of one or the other. This aspect is explicitly incorporated in our model, as discussed in the Methods Section and the Supplementary Information. Finally, 
we go beyond standard voter and opinion dynamics models by defining outcome functions modeling different electoral processes corresponding to 
different democratic countries and/or elections for different branches of government~\cite{Montesquieu}.

Quite expectedly, we find that partisan population with significant biases towards one or few parties are resilient against opinion manipulation. This is so, because
inverting the election outcome requires to change the opinion of large numbers of voters (see Supplementary Information). 
More interesting are unpartisan populations with no or only weak partisan bias, and we focus on this case here. 
Our results are that, first, 
voter populations are more resilient against external opinion manipulations (i) if they have less polarized voter opinion 
distributions and (ii) if they are more open to different opinions, i.e., when voters interact with each other regardless of their difference in opinion. Second,
applying our model to different electoral systems, we find that systems with proportional representativity are the most robust to external influences, agreeing with the recent results of Raducha et al.~\cite{Rad23}.
These results emphasize in particular the need to encourage public debates during political campaigns, to strengthen democratic processes. They further suggest to turn to systems
where representatives are attributed proportionally to the votes received by each party, at the level of electoral units corresponding to several aggregated, formerly single-representative districts. 

It should be stressed right away that our model is not meant to predict, nor to quantitatively model democratic electoral processes, but rather to extract electoral trends that can be  attributed to general social characteristics of voter populations or to specific electoral modalities. In that spirit, the model is qualitatively validated as it reproduces the observed volatility of 
the US House of Representatives elections from 2012 to 2020. We think this is quite remarkable, in particular given the notorious difficulty to validate results in computational social science~\cite{Fort2013,Burg2016,Galesic2019}.

\section*{Results}

Many different dynamical models of opinion formation in computational social science have been proposed over a remarkably long history~\cite{French1956,Har59,Abelson1967,Taylor1968,DeGroot1974,Friedkin1990,Deffuant2000,Sznajd2000,Hegselmann2002,Lorenz2007,Cas09,Hegselmann2015,Reg20}.
Most exhibit unrealistic behaviors such as consensus -- all agents end up having the same opinion -- or fragmentation -- agents have one of a handful of different, discrete opinions.
Such behaviors are not observed in real life, where it is obviously uncommon for many different people to have exactly the same opinion on a given, complex topic. 
Furthermore, even when models are free of such shortcomings, there are rather few  direct validation against real-life data and observed social phenomena~\cite{Filho1997,Filho2003,Burg2016,Galesic2019}.
Quite often, comparisons are indirect and statistical in nature~\cite{Cas09,Fort13}, an important part of the difficulty being to translate a mathematical object -- an agent's opinion vector --
into an observed sociological quantity. 
Any new model of computational social science should address these issues. Our model is presented in detail in the Methods Section and the Supplementary Information. It
successfully meets these two challenges.
The absence of consensus and fragmentation is guaranteed by the presence of a natural opinion that agents would have, were they isolated from their peers (see the Methods Section).
Validation is obtained by direct comparison between historic US House of Representatives elections and numerical simulations, taking into account different initial
partisan biases in different electoral districts, reflecting the outcome of the previous election cycle. 
This validation is discussed in the next paragraph, following which we will present our results on the robustness for different electoral systems and  
electorates with various social characteristics.

\begin{figure*}
 \centering
 \includegraphics[width=\textwidth]{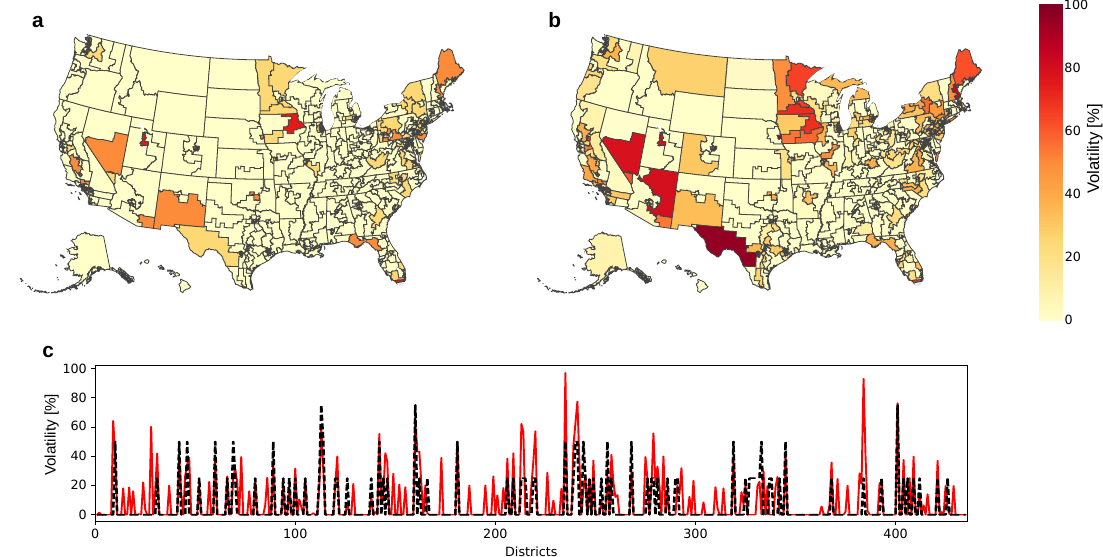}
 \caption{
 {\bf Volatility of US House of Representatives elections.}
 {\bf a} Color plot of historical percentage of party change (Republican to Democrats or vice versa) for the 435 US congressional districts over the 2012 to 2020 elections~\cite{us-election-results}. 
 {\bf b} Color plot of electoral volatility of the 435 US congressional districts, numerically calculated from our model. 
 {\bf c} Historical percentage of party change (black dashed line) vs. numerically calculated volatility (red solid line) for the 435 US congressional districts (in alphabetical order on the horizontal axis). 
 There is a Pearson correlation coefficient of 0.65 between the two sets of data.}
 \label{fig:validation_HOR}
\end{figure*}

\subsection*{Model Validation}\label{sec:validation}

Model validation requires comparison between numerically obtained data and real life, 
measurement data. For the latter, we took the US House of Representatives electoral outcomes from 2012 to 2020~\cite{us-election-results}, i.e., a complete 
election sequence in between two consecutive 
census-based redistrictings. Reproducing precise percentages of votes for the Democrats or the Republicans within districts obviously lies beyond the reach of any mathematical model.
However, trends indicated by the number of times the winning party changes from one election to the next in a given district may be captured.
In that spirit we define electoral volatility, first historically -- i.e. considering real, past elections -- 
as the fraction of times the winning party has changed in a sufficiently long sequence of elections and second numerically as the easiness with which an election can be overturned in our model.
We compare  
historical volatility of the US House of Representatives elections with numerical volatility, calculated over an ensemble of election sequences (see Supplementary Information). 
The comparison between the historical number of winning party  reversal and the numerically computed volatility is
shown in Fig.~\ref{fig:validation_HOR}. Obvious differences exist, however, the two sets of data look overall similar, which is confirmed by a Pearson correlation coefficient of 0.65, indicating a rather 
high cross-correlation between them. We conclude that our model captures the main trends and features of real elections. Given the low number of independent parameters in our model, we 
take this as a successful validation test. 

\subsection*{Bipartite System}\label{sec:2party}

Modeling an electoral process proceeds in two steps. 
In the first step, opinions are formed. In the second step, an electoral outcome is defined from these opinions. The first step 
depends on parameters that model sociological characteristics of the voter population, such as the polarization and a priori partisan biases of natural opinions, as well as
the openness of the voters to different point of views, modeled here by the interaction distance $\epsilon$ (see Methods Section). The second step models different electoral systems existing in 
different democratic countries or for different branches of government -- executive, legislative and judiciary~\cite{Montesquieu}.
To understand
which sociological aspects of a voter population influence electoral robustness and how, and which electoral system is more robust,
we would like to identify the impact of each aspect on the robustness of the outcome of an election, independently of other biases. 
We first focus on the influence of sociological characteristics of the electorate on electoral robustness (see the Methods Section).

\begin{figure}
 \centering
 \includegraphics[width=.3\textwidth]{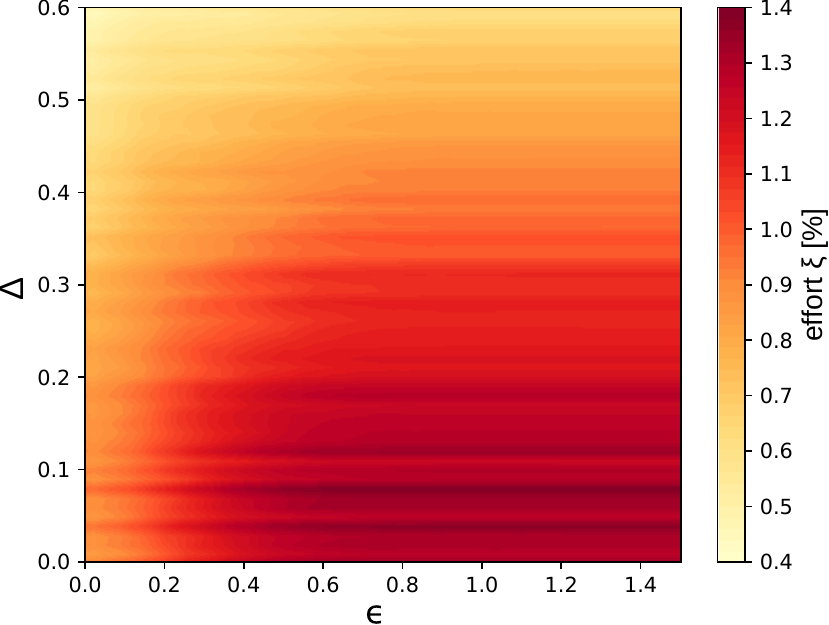}
 \caption{
 {\bf Robustness of bipartite elections vs. opinion polarization and openness in a single electoral unit.}
 Color plot of the effort needed to invert the outcome of a single-representative electoral system in a single electoral district vs. the polarization $\Delta$ of natural opinions
 and the voter interaction distance $\epsilon$. The effort is averaged over 500 realizations of natural opinions from unbiased distributions with 
 a standard deviation $\sigma=0.2$ (see the Methods Section), for an electoral district
 with $N=2001$ agents.
  }
 \label{fig:Variation_delta}
\end{figure}

Our first result is that introducing significant biases through the parameters $\mu$ and $\rho$ in Eq.~\eqref{eq:distribution} results in clear-cut election outcomes that are
very hard to invert. Our first, not unexpected result is therefore that partisan populations are quite resilient against opinion manipulation. This is a relatively trivial result, as 
significant electoral biases result in clear-cut elections with a comfortable voting margin between winning and losing parties. Reversing the electoral outcome accordingly requires
to change the opinion of large numbers of voters.  Details are discussed in Section V of the Supplementary Information.

More interesting are unbiased, or weakly biased electoral populations and we focus on them from now on.
Fig.~\ref{fig:Variation_delta} makes it clear that unpolarized (with low $\Delta$), open (with large interaction distance $\epsilon$)
voter populations are more robust to external influence. As a matter of fact, inverting the outcome of an election in a bipartite system is more efficiently achieved 
by flipping the opinion of few weakly opinionated voters -- those whose opinion vectors lie close to the neutral opinion ${\bf x}=(0.5,0.5)$. Shifting the opinion of these voters
is then easier if voter-voter interactions connect them to only few other voters. In graph theoretical vocabulary, a larger robustness is expected if weakly opinionated voters
lie in a high connectivity region of the voter-voter interaction graph.
Connectivity is higher both in the absence of
polarization, because then the density of voters close to the neutral opinion is maximal, and with long-range voter-voter interaction, which obviously connects more voters 
to one another. 
This reasoning qualitatively explains the results presented in Fig.~\ref{fig:Variation_delta}.
In the Supplementary Information, further data are presented giving the effort needed to invert the electoral outcome for biased opinion distributions (see Fig.~S6).

We next investigate the robustness of different electoral systems
for a fixed distribution of natural opinions. We compare the effort needed to change the outcome of 
the proportional representative (PR), winner-takes-all (WTA) and single representative (SR) systems, which are 
three of the main democratic electoral systems (see the Methods Section). To compare them as fairly as possible, we construct ensembles of 
synthetic countries made of between $16$ and $20$ states, each of which electing between $3$ and $15$ representatives.
In the PR and WTA systems, representatives are attributed either proportionally to the 
percentage of votes obtained by each party (PR) or all to the winning party (WTA), depending on the electoral result at the state level. 
In the SR system, only one candidate is elected in each state, independently of its population. 
To make the election more realistic, a partisan bias is introduced, which randomly favors one or the other
party in each state/district, with a difference of votes not exceeding 10\%. The bias is introduced in either a shift of the distribution of natural opinion away from zero average [the parameter $\mu$ in Eq.~\eqref{eq:distribution}], or a weight bias in 
the case of a polarized, bigaussian distribution [the parameter $\rho$ in Eq.~\eqref{eq:distribution}]. This allows us to explore different distributions of natural opinions. 
Details of the procedure are described in the Methods Section.

\begin{figure*}
 \centering
 \includegraphics[width=\textwidth]{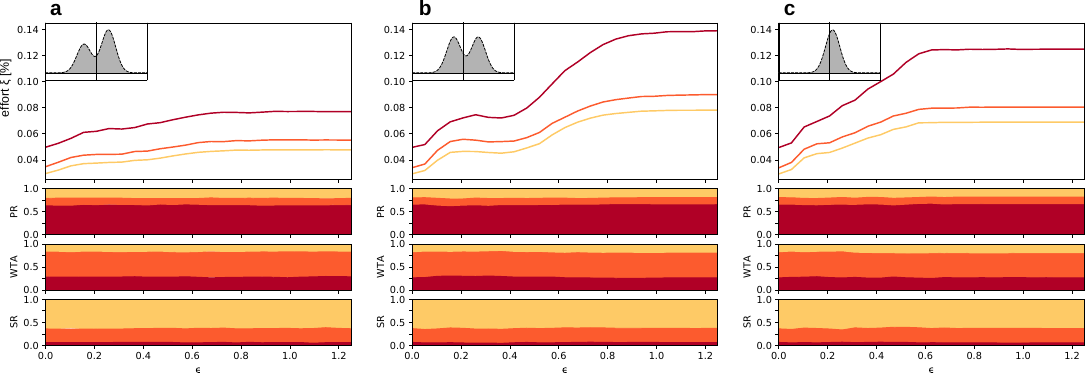}
 \caption{
 {\bf Robustness of different bipartite electoral systems vs. openness in synthetic countries.}
 Average effort to change the election outcome for different electoral systems with respect to the opinion distance over which voters interact with one another,  
 for three different types of voter population and for the proportional representative (red curves), winner-takes-all (orange) and single representative (yellow) electoral systems.
The average is calculated over an ensemble of 1500 numerically generated synthetic countries. 
 Top panels:
 {\bf a} unshifted polarized distribution with biased weight; 
 {\bf b} shifted polarized distribution with equal weights; 
 {\bf c} shifted, unpolarized distribution.
 In all cases, biases are introduced to favor one or the other party by a percentage randomly distributed between 0 and 10 \%.
 Bottom panels: probability for each system to be the most (red) the second most (orange) or the least robust (yellow) of the three systems.
 }
 \label{fig:Robustness}
\end{figure*}

Fig.~\ref{fig:Robustness} shows that the PR system is on average the most robust and the SR the least robust to external influence. This is so regardless of the distribution of natural opinions, including 
the nature of the random shift giving the initial electoral bias,  or the interaction distance between voters. Furthermore, and consistent with the results shown in Fig.~\ref{fig:Variation_delta} the average 
robustness is in all cases a monotonously increasing function of the interaction distance, regardless of the electoral system. Fig.~\ref{fig:Robustness} furthermore
shows that, beyond the average robustness, the most resilient electoral system is the PR system in  about 65 \% of the 1500 statistical realizations of synthetic countries we considered,
the WTA system in about 25 \% and the SR in less than 10 \% of those cases. In the Supplementary Information, these results are corroborated by similar data obtained  
for a model of the US House of Representatives election, calibrated as in Fig.~\ref{fig:validation_HOR} (See Fig.~S7).
Our results remarkably agree with the conclusions of Ref.~\cite{Rad23}, where the authors independently performed a similar study based on a completely different approach. 

\subsection*{Multipartite System}\label{sec:pparty}

We now turn our attention to elections with more than two parties. In this new case, robustness is quantitatively measured by the minimal effort to make the first runner-up win the election.
The Methods Section describes how we extend our bipartite system -- which belongs to the class of one-dimensional continuous opinion dynamics models~\cite{French1956,Har59,DeGroot1974,Friedkin1990} -- 
to models of interacting agents with opinion vectors of dimension larger than two. Each component of an opinion vector measures that agent's affinity with the corresponding party. 
Political parties differ from the more conceptual issues considered so far in the literature on multidimensional opinion dynamics ~\cite{Deffuant2000,Lorenz2008,Nedic2012,Friedkin2018,Stamoulas2018,Baumann2020a}
in that they are ordered from left to right. Accordingly, we need to define an opinion distance metrics that 
takes this ordering into account. In particular, the distance between the extreme left- and right-wing parties corresponds to maximally different political opinions and must therefore be larger
than the distance between any other pair of parties.  Enforcing that condition however reduces the opinion volume of left- and right-wing parties, compared to centrist parties. If uncompensated, 
this reduction of volume would increase in its turn the density of voters, making extremist parties more attractive  
under the dynamics of our model. To remove that artifactual effect, we reduce the volume occupied by all parties. 
The procedure to do so is constructed and discussed in detail in the Supplementary Information. It guarantees that there is no artifactual bias in our model.

\begin{figure}[h]
 \centering
 \includegraphics[width=.3\textwidth]{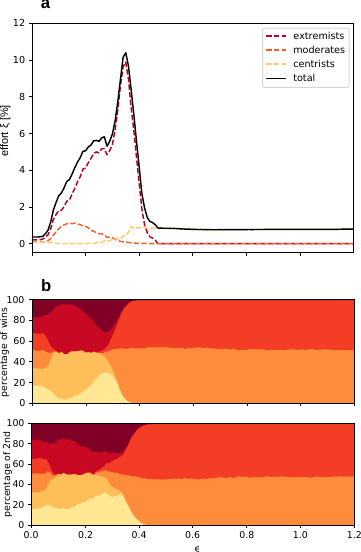}
 \caption{
 {\bf Robustness of a 6-party system vs. openness in a single electoral unit.}
 {\bf a} Effort needed to overturn a 6-party election under the single representative electoral system, in a single electoral unit (solid black curve).
 The total effort is broken down into partial efforts corresponding to overturns in favor of a centrist (dashed yellow), a moderate (dashed orange) 
 or an extremist (dashed red) party. The total effort is dominated by the effort devoted to overturns in favor of an extremist.
 {\bf b} Percentage of wins (top panel) and of second place (bottom panel) for each party in the absence of external influence.
 Different colors correspond to different parties in a left-right order when going from yellow to dark red. Larger interaction distances favor centrists parties.
 Results for different number of parties (presented in the Supplementary Information) corroborate this trend. 
 }
 \label{fig:6parties-1}
\end{figure}

In this paragraph we focus on 6-party systems.
Similar results are obtained for systems with 3 to 7 parties (see Supplementary Information, Figs.~S8 and S9), indicating 
that general features and trends of robustness do not depend on the number of parties.

We first consider an election in a single electoral district. 
Fig.~\ref{fig:6parties-1} shows the effort needed to change the election outcome in a single electoral unit with six parties (black solide line) as a function of the interaction distance $\epsilon$. 
One sees that, unlike the two-party case, the effort is here nonmonotonous in $\epsilon$. This is however an artifact of having extremist parties connected to few neighboring parties when $\epsilon$ is small,
as we next proceed to argue. 

Our models are symmetric in that the percentage of wins, of second, third aso. place finish are symmetrically distributed over parties ordered from 1 (leftmost) to 6 (rightmost). This is shown in Fig.~\ref{fig:6parties-1}{\bf b}. Fig.~\ref{fig:6parties-1}{\bf a} further breaks down the effort needed to 
change the election result when the first runner-up -- the would-be winner once sufficiently strong 
external influence is turned on -- is an extremist (party 1 or 6), a moderate (party 2 or 5) or one of the two central parties (3 and 4). It is clearly seen that, 
up to $\epsilon \simeq 0.45$, almost all the average effort needed to overturn an election corresponds to those cases where the first runner-up is an extremist. This is so, because inter-agent interactions are attractive and symmetric. Therefore,
attracting moderate voters toward extremist parties simultaneously attracts extremist voters toward moderate parties. Because extremists have such attraction only on one side of their opinion volume,
they are more easily pulled toward the center than moderate parties are toward extremes. This effect results in a much larger efforts to overturn an election in favor of an extremist party. 
Once the extremist contribution is subtracted from the effort, the latter is essentially monotonous in $\epsilon$, as in the bipartite system.  

\begin{figure*}
 \centering
 \includegraphics[width=.75\textwidth]{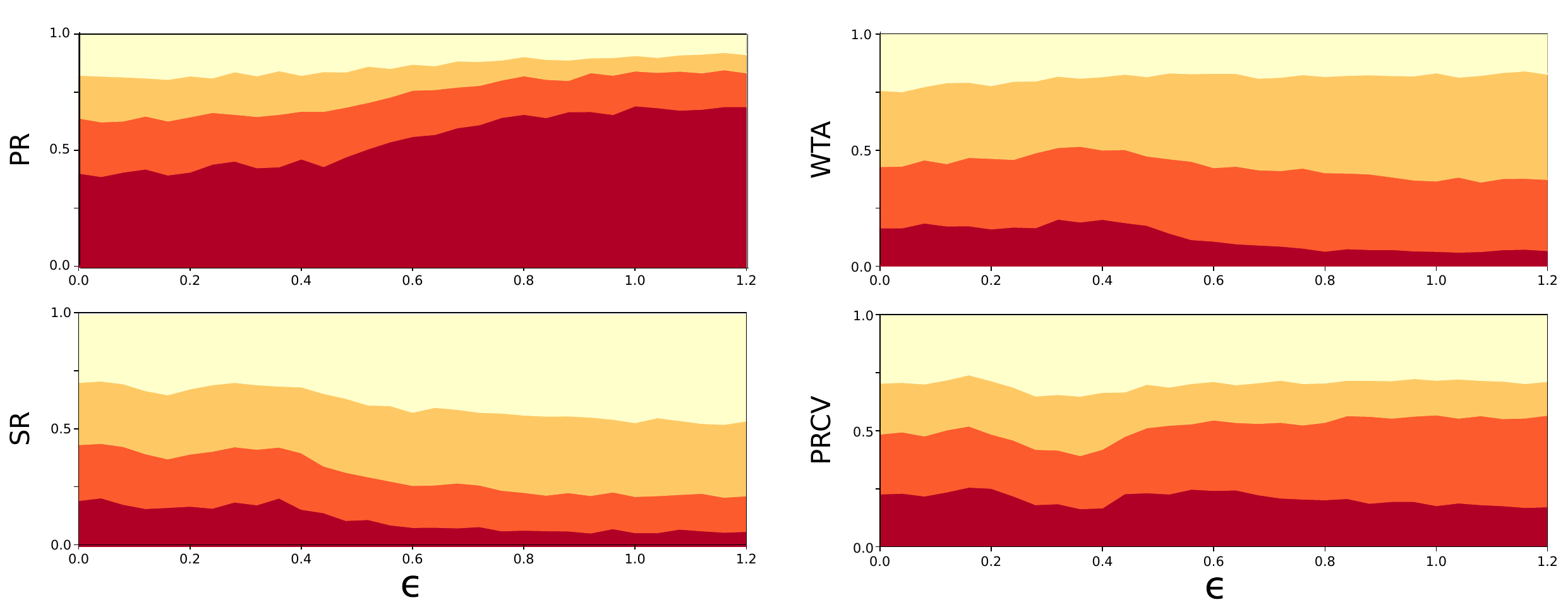}
 \caption{
 {\bf Robustness of different electoral processes for a 6-party system vs. openness in a synthetic country.}
Percentage of first- (dark red), second- (orange), third- (yellow) and fourth-rank (light yellow) robustness for the proportional representative (PR),
the winner-takes-all (WTA), the single representative (SR) and a fourth system, not existing for two-party systems, the proportional ranked choice voting (PRCV) electoral systems vs. the interaction distance $\epsilon$. Percentages are calculated over 
15000 random realizations of synthetic countries with different number of electoral units. 
 }
 \label{fig:6parties-2}
\end{figure*}

The second dominant trend observed in Fig.~\ref{fig:6parties-1} is that larger interaction distances, $\epsilon \gtrsim 0.45$, exclusively favor centrist parties, which is expected in a system
with sufficiently long-range, attractive interactions.

We finally compare different electoral systems over the same set of synthetic countries considered for the bipartite system. 
Figure~\ref{fig:6parties-2} shows the percentage of first-, second- and third-rank robustness for
the proportional representative (PR), the winner-takes-all (WTA), the single representative (SR) and the proportional ranked choice voting (PRCV) electoral systems, 
as a function of the interaction distance $\epsilon$. It clearly shows that the PR system is the most robust, most of the time,  with a
percentage of first-rank robustness starting slightly below 50 \% at small $\epsilon$ and increasing to about 65 \% with increasing $\epsilon$.
The second most robust system is the PRCV system, if judged from first-ranked robustness. Note however that it is the least robust about 30 \% of the time. 
As for the bipartite case, the WTA system is more robust  than the SR system. 

\section*{Discussion}

The process that eventually leads to an electoral result is complex. It depends on both the electorate --  in particular its polarization, its electoral bias and openness to different opinions  -- and the electoral system itself --
which translates voter opinions into an electoral outcome. Real elections cannot be reproduced under different conditions, it is therefore virtually impossible to 
disentangle the impact that each of these has on electoral processes. Well constructed models of computational social science can help identify what impact specific characteristics of voter populations
and of different electoral systems have on electoral outcomes, as well as the robustness thereof against targeted external influences. Such assessments are particularly interesting, given the recent rise of 
computational propaganda~\cite{sanovich2017a,woolley2017a,How2018a,How2018b}. Indirect electoral manipulation through the automated generation and spread of fake news is hard to stop, since it takes less time
to create bots account in social networks than to identify and neutralize them. Therefore it is highly desirable to find complementary measures, beyond social network regulation, to try and mitigate the impact
of computational propaganda on democratic elections.

This manuscript contributes to this program in several ways. To the best of our knowledge, it is among the first of its kind (with \cite{Rad23}) in computational social science to try and evaluate the robustness of different, democratic 
electoral systems. To do so, we constructed a model of opinion dynamics derived from the standard model of Taylor~\cite{Taylor1968}. We included significant additions and modifications to make
it realistic, in that it avoids reaching full consensus or fragmentation. Different voter populations have been considered, that depend on few parameters translating social traits such as voter polarization,
or their willingness to debate with voters of strongly differing opinions. These characteristics are encoded in different distributions of what we called {\it natural opinions} -- the opinion that voters would
have, were they not interacting with others, nor subjected to external influence -- and on a distance over which voters interact with one another in opinion space. We furthermore extended Taylor's model~\cite{Taylor1968} to account for the existence of more than two political parties. 
Compared to earlier constructions, ours takes into account the ordering of parties, from left to right. Last but not least, we introduced a minimal model for external influence, whose magnitude 
can be tuned until election outcomes are overturned. In this way, electoral robustness can be quantified. 

It can certainly be argued that no mathematical model can capture the whole complexity of human opinion dynamics. Instead, our goal in this manuscript has been to identify trends, i.e., how each of the few parameters
defining our model impacts the robustness of the electoral outcome. Our first result of importance has been to validate our model against 
historical data. In particular, we reproduced the volatility of the US House of Representatives elections. Such validations are rare in computational social science and this is the first important result presented in 
this manuscript. 

In this paper, we focused on unbiased or weakly biased elections -- our goal was to isolate the importance of various social and electoral characteristics, other things being equal. 
This is of course not realistic, as no electorate exist with equal distributions of natural opinions in favor of each party. We have commented that significant electoral biases strengthen
electoral results, because in their presence, more voters need to change their opinion to change the electoral outcome. A related issue of importance is that of gerrymandering, where 
electoral districts are drawn to favor one party over the others. A brief comment in the Supplementary Information shows that, at least in the case of an priori close election, with two parties
of similar strength, gerrymandering increases electoral robustness, because it introduces bias. 

In terms of the opinion landscape, we found that election outcomes are more robust in less polarized societies -- this is illustrated in Fig.~\ref{fig:Variation_delta}.
Interactions between agents have a structuring effect on the final opinions and in a more connected society, opinions become more rigid, due to the density of the interaction network. 
Increasing the interaction distance between voter's opinions naturally makes the electoral process more robust against external influence. In other words, opinions formed by interactions
with many other voters, with strongly differing opinions are harder to influence. Simultaneously, 
a similar reasoning explains that the more a society is polarized, the less it is robust against external influence. 
Indeed, polarizing an electorate increases the distance between agents in the opinion space. 
Therefore, for a limited interaction distance, there will be less interactions between agents if they are more polarized. This is especially true for weakly opinionated voters, which are rare in polarized societies,
and therefore interact with only few others. Their opinions are accordingly more easily manipulated. 

Last but not least, we investigated the robustness of different electoral systems. Focusing on the four main ones -- proportional representation (PR), 
winner-takes-all (WTA), single representation (SR) and proportional ranked choice voting (PRCV) --
we found that the PR system is generally the most robust, in remarkable agreement with the recent study by Raducha et al.~\cite{Rad23}. This is not trivial. As a matter of fact, 
on the one hand, it is easier for an influencer to gain seats in the PR system than in the SR system, because each seat in the PR system corresponds to a smaller number of votes. 
From that point of view, one would guess that the SR system is more robust than the PR system. 
On the other hand, however, the influencer needs to gain more seats to take the majority in the PR system, because in total, there are more seats available, compared to the SR system. 
To make a long story short, overturning an election in the PR system requires to flip many seats that are easy to get, while in the SR system, flipping a few seats is sufficient, each of which is however more difficult to flip.
Our results make it clear that the net result is that the PR system is the most resilient to external influences, most of the time. 
The WTA system can be seen as a middle ground between PR and SR, attributing seats in the same way as the SR system, but where the total number of seats is the same as in the PR system. 
It is not clear {\it a priori} where the WTA system would rank in terms of electoral robustness. We found that it is generally more robust than the SR, less so than the PR systems. 
The PRCV system is somehow more intricate to understand and accordingly, has a more intricate behavior. On the one hand, it is the most robust 20--25 \% of the time, which would
put it as second most robust, behind the PR system. This likely follows from the proportional attribution of seats. However the PRCV system is also the least robust about 30 \% of the time.
This indicates that, quite often, efficient strategies exist to target voter rankings to successfully modify the electoral outcome. 

Our results remain the same, regardless of the number of parties, the shape of the opinion landscape, or the  interaction distance between different voters. 
Systems with proportional representation are generally more robust than systems attributing all seats to the winner, themselves being generally more robust than single representative systems, 
irrespective of the structure of the electorate.

Altogether, these results indicate that more homogeneous, more open societies, i.e., where voters interact more with one another, and where the spectrum of opinions is smoothly represented are more robust 
against external influences. Mathematically, this is so because such societies have
a more connected voter-voter interaction graph. Democracies are strengthened when people discuss and debate more, even with people of
strongly differing opinions. Our results corroborate earlier conclusions drawn in political/social science~\cite{Barber1984,Wyatt2000}. All our results, including this conclusion, are 
independent of the number of existing political parties, However, different electoral processes behave remarkably differently under the influence of external manipulation, and we found that proportional 
representation makes for a more robust electoral process. Tradition has it in many democratic countries that a voter has a unique parliamentary representative, elected at the level of the electoral district where
that voter lives. Aggregating electoral districts, say, at the level of states/provinces, where representatives are elected proportionally would strengthen the resilience of the electoral process against 
computational propaganda -- not to mention that it would eliminate gerrymandering. 

Finally, here we focused on four of the main existing electoral systems. This work should be extended to account for other systems, including ranked-choice ballots, two-round systems and so forth. 
Works along those lines is underway. 

\section*{Methods}

\subsection{The opinion dynamics}\label{sec:od}
Upon interactions with peers, the opinion of an agent may change, i.e., move in the opinion space.
We model opinion dynamics of a group of $n$ agents with an adaptation of the model proposed by Taylor~\cite{Taylor1968} [see in particular Eq.~(3) there], which we detail here. 
 
In a democratic system with $p\ge 2$ parties, we represent the opinion of each agent as a normalized vector
on the regular $(p-1)$-simplex,
\begin{align}\label{eq:simplex}
 \Sr &= \left\{\bm{x}\in\mathbb{R}^p ~\colon~ \sum_{k=1}^p x_k = 1\, , x_k\geq 0\right\}\, , 
\end{align}
whose vertices are ${\bm v}_i=(\delta_{1i}, \delta_{2i}, ..., \delta_{pi})$, with the Kronecker symbol $\delta_{ij}=1$ if $i=j$ and $\delta_{ij}=0$ otherwise.
The use of the 1-norm in equation~\eqref{eq:simplex} reflects the fact that each voter has a global opinion.
The $k$-th component of $\bm{x}\in \Sr$ represents the affinity of that agent with the $k^{\rm th}$ party. Accordingly,  
the largest component determines the party the agent will vote for.
The opinion space will be refined in the next section in order to account for party ordering without introducing artificial, mathematical biases in favor of one party or another. 

We attribute a natural opinion $\bm{x}_i^{(0)} \in \Sr$ to each agent $i$, and gather all natural opinions in the natural opinion matrix $X^{(0)}\in\mathbb{R}^{n\times p}$ whose $i$-th row is $\bm{x}_i^{(0)}$. 
The natural opinions of an agent is  the actual opinion they would have in the absence of agent-agent interaction.
Distance between natural opinions determine these interactions.
In the spirit of the Deffuant
~\cite{Deffuant2000} and Hegselmann-Krause~\cite{Hegselmann2002} models of opinion dynamics, interactions exist only between pairs of voters with opinion vectors within a certain distance $\epsilon$ 
in the agent opinion space. 
This defines the adjacency matrix $A$ of the interaction graph,
\begin{align}
 A_{ij} &= \left\{
 \begin{array}{ll}
  1\, , & \text{if } \|{\bm x}_{i}^{(0)} - {\bm x}_{^j}^{(0)}\|_1 < \epsilon\, , \\
  0\, , & \text{otherwise,}
 \end{array}
 \right.
\end{align}
where the interaction distance parameter $\epsilon$ is sometimes called the \emph{confidence bound}~\cite{Deffuant2000,Hegselmann2002}.

Finally, each agent forms their own opinion by balancing their neighbors opinions, following a consensus algorithm, and their own natural opinion. 
The dynamics of agent $i$'s opinion towards party $q$ is given by 
\begin{align}\label{eq:dyn-noinf}
 \dot{X}_{iq} &= d_i^{-1}\sum_{j=1}^n A_{ij}(X_{jq} - X_{iq}) + (X^{(0)}_{iq} - X_{iq})\, ,
\end{align}
where we introduced the degree 
$d_i=\sum_{k=1}^n A_{ik}$ of $i$ in the interaction graph to normalize the total influence of other agents on agent $i$. In words, 
the influence that a voter has on a given voter $i$ is inversely proportional to -- is diluted by -- the total number of voters interacting with voter $i$.

The final steps are to define the diagonal degree matrix 
$D = {\rm diag}(d_i)$ and the Laplacian matrix $ L = D-A$, to write 
equation~\eqref{eq:dyn-noinf} in vector form
\begin{align}\label{eq:vec_dyn-noinf}
 \dot{X} &= {X}^{(0)} - \left[ D^{-1} L + I_n\right]X\, ,
\end{align}
with the $n$-dimensional identity matrix $I_n$.

Equation~\eqref{eq:vec_dyn-noinf} is a linear system and the matrix $D^{-1}L + I_n$ has strictly positive eigenvalues (see Supplementary Information). 
Therefore, the system is globally exponentially stable and converges toward the equilibrium 
\begin{align}\label{eq:equilibrium}
 X^* &= \left[D^{-1}L + I_n\right]^{-1}{X}^{(0)}\, .
\end{align}
In our model, that equilibrium defines the final opinion of the agents. In other words, we assume that by election day, the dynamic process has converged. 

We model the external influence applied by computational propaganda as a slight shift in the agents' natural opinions. 
Accordingly, we modify the opinion dynamics of equation~\eqref{eq:vec_dyn-noinf} as 
\begin{align}\label{eq:vec_dyn}
 \dot{X} &= {X}^{(0)} + W - \left[ D^{-1} L + I_n\right]X\, ,
\end{align}
where $W\in\mathbb{R}^{n\times p}$ encodes the external influence and satisfies 
$ \sum_{j=1}^n W_{ij} = 0 $ and $-X^{(0)}_{ij} \leq W_{ij} \leq 1 - X^{(0)}_{ij}$
to guarantee that opinions remain normalized under the dynamics defined by equation~\eqref{eq:vec_dyn}, i.e. that they stay 
on the $(p-1)$-simplex.

A measure of the \emph{robustness} of an election outcome is the minimal amount of external influence that needs to be applied in order to change the outcome. 
In our model, we introduce influence step by step, starting from the uninfluenced sytems ($W_0={\bm 0}_{n\times p}$) and adding small increments of influence until the election outcome changes. 
We have found that the most efficient influence strategy is the following one:
\begin{itemize}
 \item the external influence attempts to help the first runner-up of the election, which has the most chance to steal the elections from the winner;
 \item agents whose natural opinion is close to the boundary to the first runner-up are targeted first;
 \item when multiple electoral units are involved, external influence targets first the electoral units with the lowest majority (see the Supplementary Information for further details). 
\end{itemize}
In practice, when we want to target the $i^{\rm th}$ agent, and push their opinion towards the $q^{\rm th}$ party (the first runner-up), we place a vector
\begin{align}
 {\bm w}_k &= (w_1,...,w_p) \perp \bm{1}_p\, ,
\end{align}
on the $i^{\rm th}$-th row of the external influence matrix $W$. The components satisfy $\sum w_i=0$, with $w_q>0$ and $w_{q'}<0$ for $q'\neq q$. 

If the outcome of the election in electoral unit $u$ changes after $K_u$ increments, we define the \emph{effort} needed to change the election outcome as the proportion of agents influenced, 
\begin{align}
 \xi_u &= \frac{K_u}{n_u}\, ,
\end{align}  
where the subscript $u$ indexes the electoral units. 
When multiple electoral units are involved, the total effort is 
\begin{align}
 \xi_{\rm tot} &= \sum_u \frac{n_u}{N}\xi_u\, ,
\end{align}
where $N$ is the total number of agents in the whole country.

\subsection{The opinion space}\label{sec:opspace}
\begin{figure}
 \centering
 \includegraphics[width=.35\textwidth]{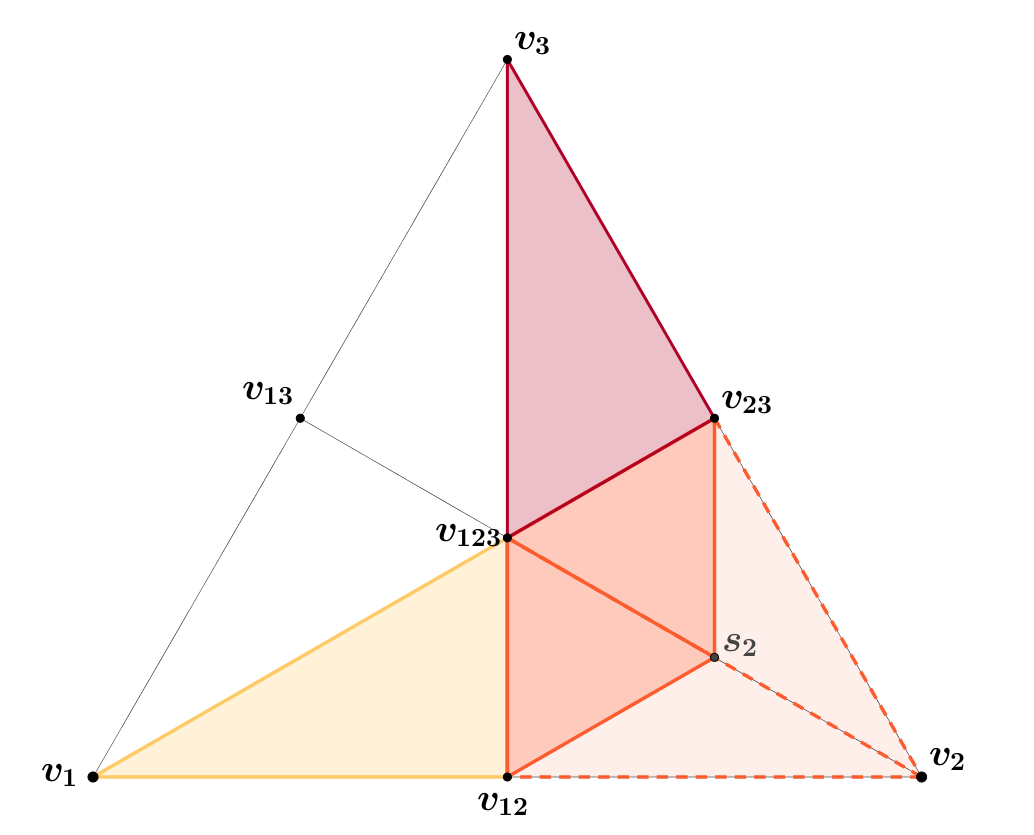}
 \caption{
 {\bf Sketch of the opinion space.}
 Admissible opinion domain in the $2$-simplex (i.e., for $p=3$ parties). 
 The vertices ${\bm v}_1$, ${\bm v}_2$, and ${\bm v}_3$ represent the pure opinions of each party respectively. 
 The shaded orange area is removed in order to balance the volume allocated to each party in the simplex, by sliding vertex ${\bm v}_2$ towards the barycenter ${\bm c} = {\bm v}_{123}$ to ${\bm s}_2$. 
 }
 \label{fig:simplex}
\end{figure}

The $k$-th component of the opinion vector ${\bm x}_i\in\Sr$ represents the degree of agreement of agent $i$ towards party $k$. 
The three polygons $({\bm v}_1,{\bm v}_{12},{\bm v}_{123},{\bm v}_{13})$, $({\bm v}_2,{\bm v}_{23},{\bm v}_{123},{\bm v}_{12})$, and $({\bm v}_3,{\bm v}_{13},{\bm v}_{123},{\bm v}_{23})$ in Fig.~\ref{fig:simplex} delineate the domains of the $2$-simplex where agents would vote for party $1$, $2$, and $3$ respectively. 

Without loss of generality, we can assume that the indexing of the parties (from $1$ to $p$) matches the left-right political spectrum, i.e., party $1$ is the left-most party and party $p$ is the right-most one. 
In order to guarantee coherence of each agent's opinions, we do not allow opinion vectors where two parties that are far away on the left-right spectrum have a large value while the parties in between have a low value. 
We need the opinion vector to satisfy some form of monotonicity. 
We say that an opinion vector $\bm{x}\in \Sr$ is admissible only if its components increase until the largest one and then decrease.
Mathematically, if $x_m = \max_i x_i$, we impose
\begin{align}\label{eq:ordering}
 x_i &\leq x_j\, ,~ \forall i\leq j\leq m\, , & x_i &\leq x_j\, ,~ \forall i\geq j\geq m\, ,
\end{align}
reducing the domain of allowed opinion. 
For instance, for $p=3$ parties, equation~\eqref{eq:ordering} cuts out half of the areas associated to parties $1$ and $3$, namely, the triangles $({\bm v}_1,{\bm v}_{13},{\bm v}_{123})$ and $({\bm v}_3,{\bm v}_{13},{\bm v}_{123})$, in Fig.~\ref{fig:simplex}. 

Furthermore, in order to isolate the impact of opinions and electoral systems on the robustness of elections outcomes, we focus our analysis on societies where all the parties have a similar level of approbation in the population. 
Therefore, we need to balance the volume $V_i$ allocated to each party $i\in\{1,...,p\}$ in the admissible opinion space. 
After the ordering imposed by equation~\eqref{eq:ordering}, one realizes that the left-most and right-most parties have the smallest volume in the opinion space. 
Therefore, we need to reduce the volume $V_i$ of the other parties $i\in\{2,...,p-1\}$ in order to match the volumes of parties $1$ and $p$. 

We reduce the volume of the \emph{moderates} (i.e., non-extremists) by sliding the vertex of the simplex defining their area ($\bm{v}_i$) towards the barycenter $\bm{c}$ of the simplex, reaching the new vertex $\bm{s}_i$ such that $V_i = V_1$ (see Fig.~\ref{fig:simplex}). 
In the Supplementary Information, we show that the position of $\bm{s}_i$ can be determined in closed form and rather elegantly as
\begin{align}
 \bm{s}_i &= \gamma_i \bm{c} + (1-\gamma_i)\bm{v}_i\, , \;\;\;\;\; \;\;\gamma_i^{-1}  = \binom{p-1}{i-1} \, .
\end{align}

\emph{A priori}, it is not obvious (especially for $p\geq 3$) that the ratio of displacement of the vertex $\bm{s}_i$ along the segment $[\bm{v}_i,\bm{c}]$ translates into the same ratio of volume reduction for the opinion subdomain. 
The proof of this fact (see the Supplementary Information) relies on an appropriate decomposition of the opinion space into sub-simplices. 
We show that the sliding of the vertex ${\bm s}_i$ towards ${\bm c}$ only scales the height of sub-simplices, without affecting their base, resulting in the same scaling in the volume of each sub-simplex. 

We will refer to the final opinion space (after ordering and reduction) as $\Cr\subset \Sr$, shown as the colored area in Fig.~\ref{fig:simplex}. 

\subsection{Natural opinions}\label{sec:natural_opinion}

Electoral populations are defined by two main social characteristic, which are their openness, modeled by the opinion distance over which two voters interact
(the interaction distance $\epsilon$) and the distribution of the voter's natural opinions. For the latter, we consider two cases separately, which are (i) the 
bipartite case and (ii) the multipartite case with more than two parties. 

For case (i), we considered distributions of natural opinions modeled by their bias towards one party, their polarization, and their width of represented opinions.
Following Refs.~\cite{Yang2020, Axelrod2021} we consider bigaussian distributions of one-dimensional opinions,
\begin{widetext}
\begin{align}\label{eq:distribution}
 f(x) &= \frac{\rho}{\sigma\sqrt{2\pi}} \exp[-\frac{(x - \mu + \frac{\Delta}{2})^2}{2\sigma^2}] +\frac{1-\rho}{\sigma\sqrt{2\pi}} \exp[-\frac{(x - \mu - \frac{\Delta}{2})^2}{2\sigma^2}]\, ,
\end{align}
\end{widetext}
with opinions restricted to $x \in [-1,1]$. Voter 
polarization is given by $\Delta$, while biases can be introduced either via differences in weight $\rho$ between the two gaussians or by the shift  
$\mu$. In this paper, the width of the distributions is fixed at $\sigma=0.2$, without loss of generality. For the sake of consistency with multipartite 
cases, our bipartite model is constructed in a space of two-dimensional voting agents, on which the interval $x \in [-1,1]$ can bijectively be mapped on,
via an affine, dynamics-preserving mapping, $x \mapsto \left( (1+x)/2,(1-x)/2 \right)$.

We may further consider polarization into more than two peaks, which naturally leads to case (ii) of multipartite systems. In that case, we did not consider
polarization beyond the existence of parties and instead generated uniformly distributed natural opinions as described in Sec.~\ref{sec:opspace}. 
We provide a random opinion generator online~\cite{repo}.

\subsection{Electoral systems}\label{sec:electoral-syst}
Once the actual opinion of each agent is known, we need to translate the set of final opinions into an election result. 
Each agent naturally votes for the party with which they agree the most, i.e., corresponding to the largest component in their opinion vector. 
Votes are then aggregated according to a chosen electoral system, which we would like to compare. 

We first define three electoral units corresponding to three different levels of aggregation:
\begin{description}
 \item[Country] In almost all electoral instances, 
 the largest possible electoral unit is a country. A notable exception are parliamentary elections in the European Union. 
 Examples of elections realized at the country level include presidential elections in Chile, France, Ireland, Mexico and several other countries.
 
 \item[State] 
 Most modern countries are partitioned into states -- other denominations exist such as provinces, cantons, departments, regions aso.
Parliamentary elections realized at the state level include elections to the US Senate, as well as National and State Council elections in Switzerland. 
The US presidential elections are a special case of elections whose outcome is determined by aggregating elections at the state level (except in Maine and Nebraska). 
 
 \item[District] 
 States are often subdivided into districts. In many instances these entities are organized in such a way that they all have more or less the same population.
In parliamentary elections, each electoral district then elects its own, single representative. Examples thereof include the UK general elections, 
 the US House of Representatives elections.

\end{description}

Depending on the considered electoral process, results are gathered at one or the other level, and, if applicable, results in different units in that level are combined to give the outcome, i.e. the winning party.

There are different ways to elect representatives in an electoral unit. In this manuscript, we focus on the four main ones, which are:

\begin{description}
 \item[Proportional representation (PR)]
 An electoral unit -- for instance a state -- elects more than one representative. Seats are attributed to parties proportionally to the number of votes they received. 
The members of the lower chamber of Switzerland's parliament (the National Council) are elected proportionally in each canton. 
  
 \item[Single representative (SR)]
 The electoral units here are electoral districts with a single representative. They can be elected in a single-round, simple majority election (US House of Representatives elections) 
 or in a two-round election (French National Assembly elections). 
 
 \item[Winner takes all (WTA)]
 All seats associated with an electoral unit are attributed to the party with the most votes. This system reduces to SR if there is a single seat attributed in the electoral unit. 
The US Electoral College appointing the president and vice president of the US is elected by a WTA process in all but two states (Maine and Nebraska). 

\item[Proportional Ranked Choice Voting (PRCV)] 
Electoral units are the same as in the PR system. 
Voters in each unit rank the candidates in order of preference. All candidates that receive a voting ratio above a certain quota are elected. Usually the 
quota is fixed around $1/(m+1)$ with the number $m$ of seats in the considered electoral unit. Next, surplus votes above quota are re-distributed to not-yet-elected 
second-choice
candidates. If new candidates exceed the quota, they are elected and their surplus votes are again re-distributed. 
If no candidate exceeds the quota, the candidate with the least vote is eliminated and their votes redistributed according to second preferences. The process
is iterated until $m$ seats are attributed. 
\end{description}

\section*{Data availability}
All data were either extracted from public sources~\cite{us-election-results} or generated by the code available online~\cite{repo}.

\section*{Code availability}
An implementation of the model, opinion generator, and of the influence process is available online~\cite{repo}.

\section*{Acknowledgments}
We thank the Swiss National Science Foundation for financial support under 
grant number P400P2\_194359 (RD) and 200020\_182050 (GMG and PJ).

\section*{Author contributions}
RD and PJ designed the research.
GMG performed the numerical simulations.
All authors participated in the construction of the theoretical model, the results discussion, and the writing of the manuscript. 

\section*{Competing interests}
The authors declare no competing financial interests.

\setcounter {figure} {0}
\setcounter {equation} {0}
\renewcommand{\thefigure}{S\arabic{figure}}
\renewcommand{\theequation}{S\arabic{equation}}

\newpage

~

\newpage

\onecolumngrid 

\appendix

\begin{center}
{\LARGE Supplementary Information}
\end{center}

\section{Opinion volume balancing}
For a system with $p$ parties, by definition [Eq.~(1)], the opinion space is a $(p-1)$-dimensional subset of $\mathbb{R}^p$. 
Therefore, for simplicity's sake, we embed the regular $(p-1)$-simplex $\Sr$ in $\mathbb{R}^{p-1}$.

Let ${\bm v}_1,...,{\bm v}_{p}\in\mathbb{R}^{p-1}$ be $p$ points 
and let 
\begin{align}
 S &\coloneqq \conv({\bm v}_1,...,{\bm v}_{p})\, ,
\end{align}
be the \textit{$(p-1)$-simplex} $S\subset\mathbb{R}^{p-1}$ with \textit{vertices} ${\bm v}_1$, ..., ${\bm v}_{p}$. 
The convex hull of $p-1$ vertices of~$S$ is a $(p-2)$-simplex, called a \textit{facet} of $S$. 
The facet of $S$ not containing the vertex ${\bm v}_i$ is denoted by $F_i$ (it is the facet opposite to ${\bm v}_i$ in $S$), that is
\begin{align}
 F_i &\coloneqq \conv({\bm v}_1,...,{\bm v}_{i-1},{\bm v}_{i+1},...,{\bm v}_{p})\, , &  i &\in \{1,...,p\}\, .
\end{align}
The angle between the facets $F_i$ and $F_j$ is called the \textit{dihedral angle} $\alpha_{ij}\in [0,\pi]$ of $S$.

The simplex $S\subset\mathbb{R}^{p-1}$ is said to be \textit{regular} if $\alpha_{ij}=\alpha_{kl}=:\alpha_{\text{reg}}$ for all $1\leq i,j,k,l\leq p$. 
It can be shown that $\alpha_\text{reg}=\text{arccos}\frac{1}{p-1}$~\cite[Section 7.9]{Cox73}.
For $1\leq q\leq p$, a $(q-1)$-simplex $S\subset\mathbb{R}^{p-1}$ is said to be a $(q-1)$-\textit{orthoscheme} if $\alpha_{ij}=\frac{\pi}{2}$ if $|i-j|>1$. 
For instance, a $2$-orthoscheme is a right triangle, and the facets of a $(q-1)$-orthoscheme are $(q-2)$-orthoschemes.\\

\begin{figure*}
 \centering
 \includegraphics[width=.45\textwidth]{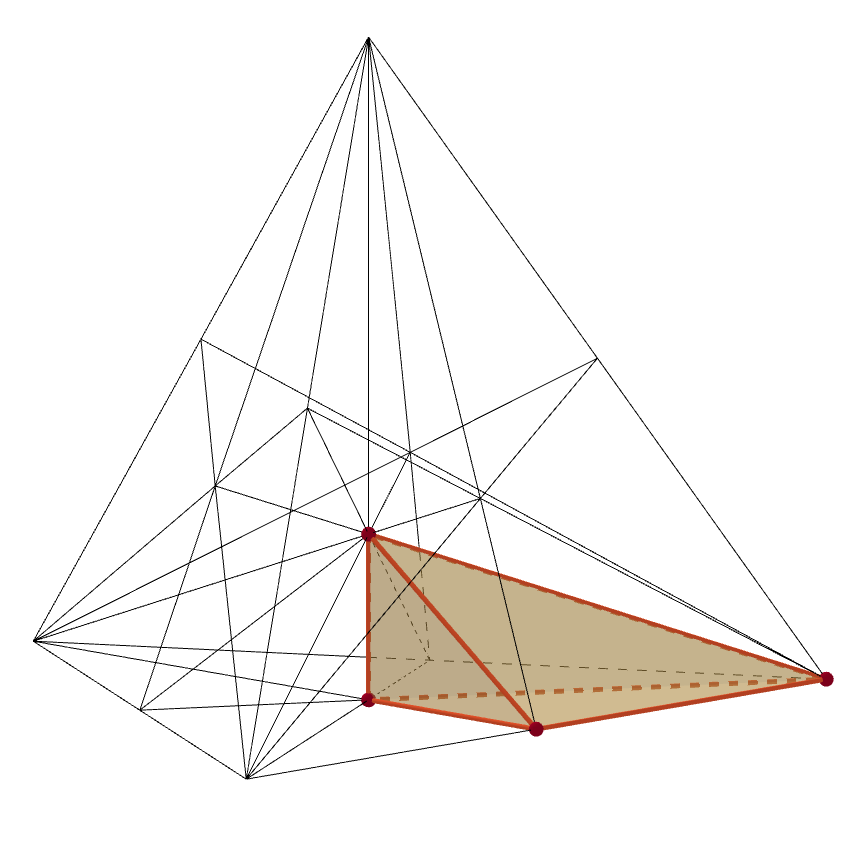}
 \caption{The regular 3-simplex $\Sr$ and one of its 24 isometric fundamental orthoschemes $O$. 
 Notice that here $p=4$.}
 \label{fig:orthoschemes}
\end{figure*}

The regular $(p-1)$-simplex $\Sr$ can be dissected into $p!$ isometric $(p-1)$-orthoschemes by means of the recursive \textit{barycentric decomposition} as follows (see Fig.~\ref{fig:orthoschemes}).
For $1\leq k\leq p$, denote by ${\bm v}_{1,...,k}$ the barycenter of the convex hull $\conv({\bm v}_1,...,{\bm v}_k)$. 
Because $\Sr$ is regular (so that all its $d$-dimensional faces are regular $d$-simplices themselves, $0\leq d\leq p-1$), ${\bm v}_{1,...,k}$ is the center of the $(k-1)$-sphere inscribed in the convex hull of ${\bm v}_1,...,{\bm v}_k$ (and of the $(k-1)$-sphere circumscribed around the convex hull of ${\bm v}_1,...,{\bm v}_k$, etc.). 
For instance, ${\bm v}_{1,2}$ is the midpoint of the edge $[{\bm v}_1,{\bm v}_2]$ of $\Sr$, ${\bm v}_{1,2,3}$ is the center of the regular triangle with vertices ${\bm v}_1,{\bm v}_2,{\bm v}_3$ (a $2$-dimensional face of $\Sr$), and ${\bm v}_{1,...,p}$ is the center of $\Sr$.
Then, the convex hull 
\begin{align}
 O_{1,...p} &\coloneqq \conv({\bm v}_1, ..., {\bm v}_{1,...,p})\, ,
\end{align}
is a $(p-1)$-orthoscheme whose non-right dihedral angles are $\frac{\pi}{3}$,...,$\frac{\pi}{3}$, $\frac{\alpha_\text{reg}}{2}$.

The procedure described above can be performed for any permutation of $\{1,...,p\}$. 
Let $\mathcal{S}_{p}$ denote the group of all permutations of $p$ objects, and let $(i_1,...,i_{p})\in\mathcal{S}_{p}$ be such a permutation. 
Then, the corresponding orthoscheme $O_{i_1,...,i_{p}}$ is given by
\begin{align}
 O_{i_1,...,i_{p}} &= \conv({\bm v}_{i_1},{\bm v}_{i_1,i_2}, ...,{\bm v}_{i_1,...,i_{p}})\, .
\end{align}
Because $\Sr$ is regular, $O_{i_1,...,i_{p}}$ is isometric to $O_{1,...,p}$ for any permutation $(i_1,...,i_{p})\in\mathcal{S}_{p}$.
This leads to a natural bijection between the set of isometric $(p-1)$-orthoschemes whose union is $\Sr$ and the set $\mathcal{S}_{p}$.

Without loss of generality, let us consider the orthoscheme $O\coloneqq O_{1,...,p}$. 
Denote by $F_{1,...,k}$ the facet of $O$ opposite to $v_{1,...,k}$. 
Now move the vertex ${\bm v}_1$ along the edge $[{\bm v}_1,{\bm v}_{1,...,p}]$ towards the vertex ${\bm v}_{1,...,p}$, to a point ${\bm s}\in[{\bm v}_1,{\bm v}_{1,...,p}]$.
Let $T$ be the simplex given by
\begin{align}
 T &\coloneqq \conv({\bm s},{\bm v}_{1,2},...,{\bm v}_{1,...,n+1})\, .
\end{align}
For a given proportionality coefficient $\kappa\in[0,1]$, we want to know where to set ${\bm s}={\bm s}(\kappa)$ on $[{\bm v}_1,{\bm v}_{1,...,p}]$ so that 
\begin{align}\label{eq:prop_vol_cond}
 \vol_{p-1}(T) &= \kappa\cdot\vol_{p-1}(O)\, .
\end{align}
It is clear that $\kappa=0\Leftrightarrow {\bm s}={\bm v}_{1,...,p}$ and that $\kappa=1\Leftrightarrow {\bm s}={\bm v}_1$.

\begin{figure*}
 \centering
 \includegraphics[width=.5\textwidth]{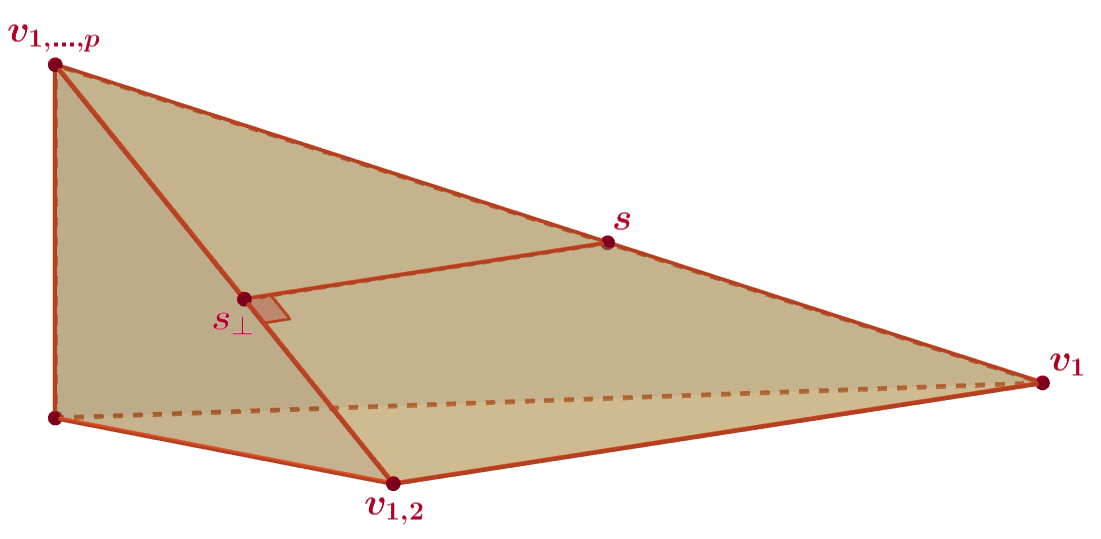}
 ~
 \includegraphics[width=.4\textwidth]{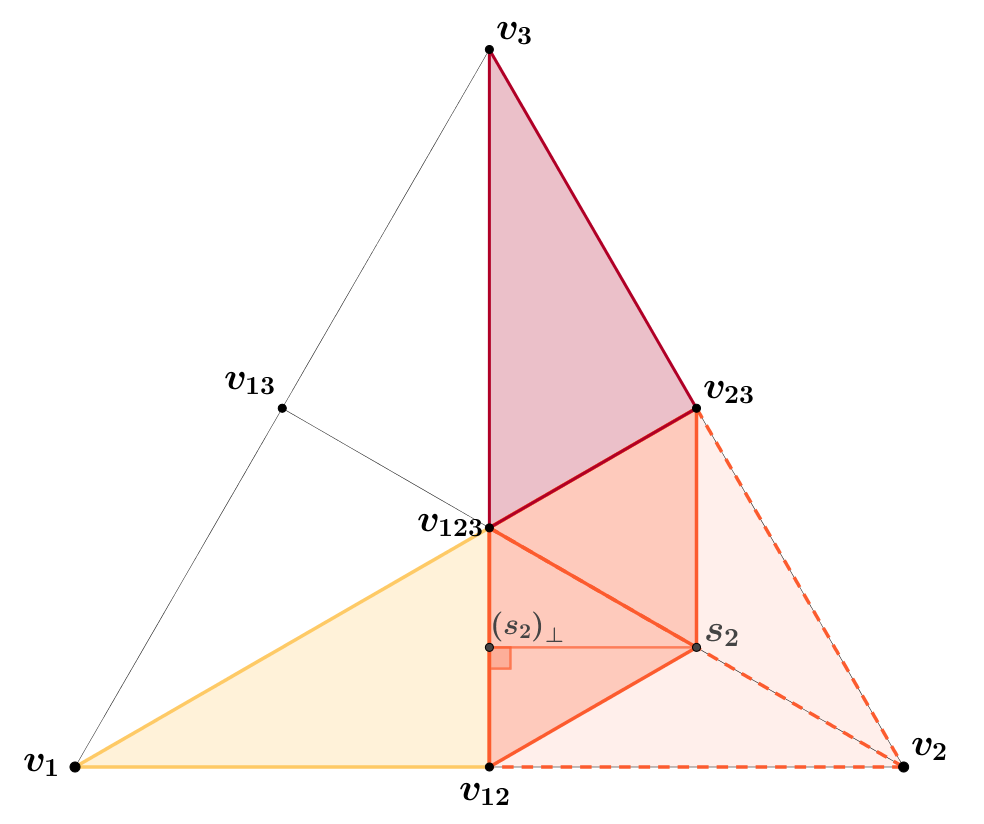}
 \caption{Left: Vertex ${\bm s}$ in the segment $[{\bm v}_1,{\bm v}_{1,...,p}]$ and its projection on the facet $F_1$.
 Right: Explicit realization of the construction of the left panel for $p=3$. }
 \label{fig:ortho-slide}
\end{figure*}

Denote by ${\bm s}_\perp$ the orthogonal projection of ${\bm s}$ on $F_1$, that is, ${\bm s}_\perp \in F_1$ and $[{\bm s},{\bm s}_\perp] \perp F_1$.
The edge $[{\bm v}_1,{\bm v}_{1,2}]$ of $O$ is orthogonal to $F_1$, since
\begin{align}
 [{\bm v}_1,{\bm v}_{1,2}] &= \bigcap_{k=3}^{p} F_{1,2,...,k}\, ,
\end{align}
and since $F_{1,...,k}$ is orthogonal to $F_1$ for $3\leq k\leq p$ (because $O$ is an orthoscheme).
See Fig.~\ref{fig:ortho-slide} for an illustration of these observations. 

Let $F_s$ be the facet of $T$ opposite to ${\bm s}$.
Because the segments $[{\bm v}_1,{\bm v}_{1,2}]$ and $[{\bm s},{\bm s}_\perp]$ are orthogonal to $F_1$ and $F_s$ respectively, the volumes of $O$ and $T$ are given by
\begin{align}
 \vol_{p-1}(O) &= \frac{1}{p-1}\cdot \dist({\bm v}_1,{\bm v}_{1,2})\cdot \vol_{p-2}(F_1)\, , 
\end{align}
and
\begin{align}
 \vol_{p-1}(T) &= \frac{1}{p-1}\cdot \dist({\bm s},{\bm s}_\perp)\cdot \vol_{p-2}(F_s)\, .
\end{align}
Now observe that
\begin{align}
 F_1 & =\conv({\bm v}_{1,2},...,{\bm v}_{1,...,p}) = F_s\, ,
\end{align}
so that the quotient of the volumes of $O$ and $T$ is given by
\begin{align}
 \frac{\vol_{p-1}(T)}{\vol_{p-1}(O)} &= \frac{\dist({\bm s},{\bm s}_\perp)}{\dist({\bm v}_1,{\bm v}_{1,2})}\, ,
\end{align}
so that Condition \eqref{eq:prop_vol_cond} is equivalent to
\begin{align}\label{eq:dist_prop_cond}
 \dist({\bm s},{\bm s}_\perp) &= \kappa\cdot\dist({\bm v}_1,{\bm v}_{1,2})\, .
\end{align}

Let us now investigate further the relative positions of the segments $[{\bm v}_1,{\bm v}_{1,2}]$ and $[{\bm s},{\bm s}_\perp]$. 
Since $\dim(F_1)=p-2$ (because $F_1$ is the convex hull of $p-1$ vertices) and since $[{\bm v}_1,{\bm v}_{1,2}]\perp F_1$ (as stated above), it follows that the segments $[{\bm s},{\bm s}_\perp]$ and $[{\bm v}_1,{\bm v}_{1,2}]$ are parallel.
Moreover, by construction, the points ${\bm v}_1$, ${\bm v}_{1,2}$ and ${\bm s}$ all belong to the $2$-face
\begin{align}
 \triangle &\coloneqq \conv({\bm v}_1,{\bm v}_{1,2},{\bm v}_{1,...,p})\, ,
\end{align}
of $O$. 
Hence, because $[{\bm s},{\bm s}_\perp]\,//\,[{\bm v}_1,{\bm v}_{1,2}]$, it follows that ${\bm s}_\perp$ is contained in $\triangle$ as well.
This in turn implies that the point ${\bm s}_\perp$ is on the segment $[{\bm v}_{1,2},{\bm v}_{1,...,p}]$, since it is an edge of $F_1$. 
Let
\begin{align}
 f &\coloneqq \conv({\bm s},{\bm s}_\perp,{\bm v}_{1,...,p})\, ,
\end{align}
denote the $2$-face of $T$ with vertices ${\bm s}$, ${\bm s}_\perp$ and ${\bm v}_{1,...,p}$.
From the discussion above we deduce that $f\subset\triangle$, and that both are right-angled triangles ($f$ at ${\bm s}_\perp$, and $\triangle$ at ${\bm v}_{1,2}$) sharing the vertex ${\bm v}_{1,...,p}$. 

These observations allow us to deduce that Condition \eqref{eq:dist_prop_cond} is equivalent to
\begin{align}
 \dist({\bm s},{\bm v}_{1,...,p}) &= \kappa\cdot\dist({\bm v}_1,{\bm v}_{1,...,p})\, ,
\end{align}
leading us to the \textit{a priori} somewhat intuitive (or seemingly too nice to be true) fact:

\begin{prop}\label{prop:vol}
In order to reduce the volume of $O$ to a proportion $\kappa\in[0,1]$, the vertex ${\bm v}_1$ has to be moved towards ${\bm v}_{1,...,p}$ to the point ${\bm s}\in[{\bm v}_1,{\bm v}_{1,...,p}]$ dividing the edge $[{\bm v}_1,{\bm v}_{1,...,p}]$ under the same proportion $\kappa$.
\end{prop}

Now that we have a way to control the volume of an orthoscheme while keeping one of its facets intact, we turn to the main question of interest for the opinion generation problem.

\begin{figure*}
 \centering
 \includegraphics[width=.43\textwidth]{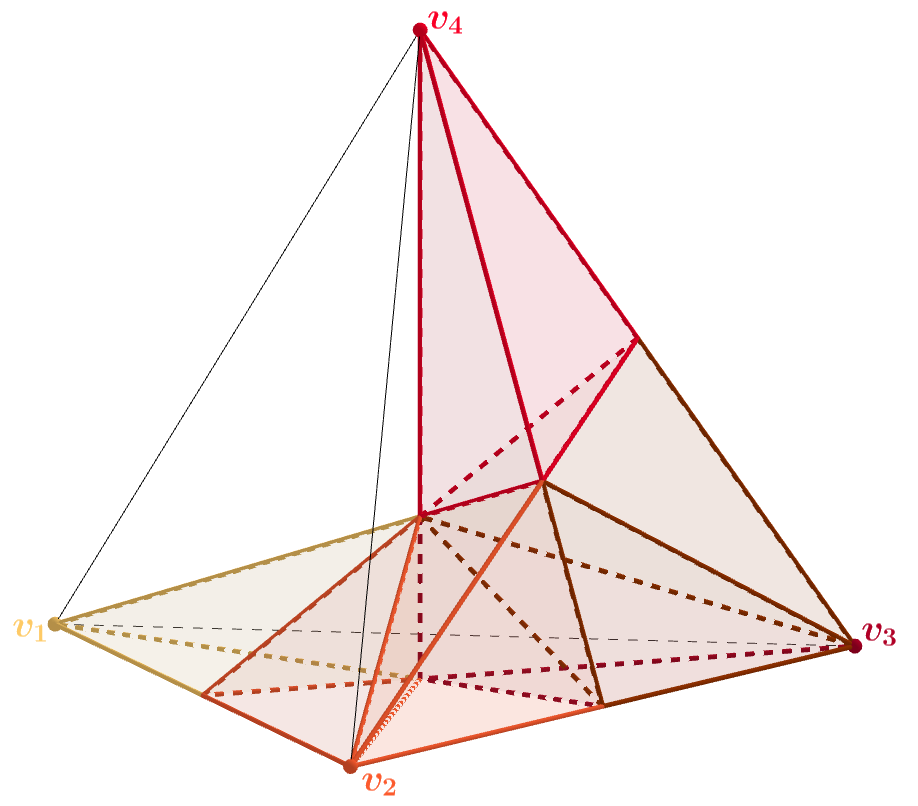}
 \caption{Illustration of the set $C$ of admissible opinions for $p=4$. }
 \label{fig:admissible-op}
\end{figure*}

Without loss of generality, label the parties from $1$ to $p$ to match the left-right political spectrum. 
In order for an opinion to be consistent, we suppose that party $i$ is the most favored (for some $i\in\{1,...,p\}$), and that the adhesion to party $j$ is smaller than (respectively, greater than) the adhesion to party $j+1$ if $j\in\{1,...,i-1\}$ (respectively, $j\in\{i,...,p\}$). 
These constraints reflect the assumption that as parties are further from the agent's favorite party on the political spectrum, the agent's adhesion towards them decreases.
Let ${\bm x}\in \Sr$ be a point representing the opinion of the agent as follows. 
The agent favors party $j$ over party $k$ if and only if $\dist({\bm x},{\bm v}_j)<\dist({\bm x},{\bm v}_k)$. 
The set $C$ of all points ${\bm x}\in \Sr$ representing admissible opinions (in the sense described above) is given by the non-convex polyhedron (see Fig.~\ref{fig:admissible-op})
\begin{align}\label{def:consistency_polyhedron}
 C &= \left\{{\bm x} \in \Sr\,\middle|\,
 \begin{array}{c}
  \text{There is an }i\in\{1,...,n+1\}\text{ such that} \\
  \dist({\bm x},{\bm v}_j) \geq \dist({\bm x},{\bm v}_{j+1}) \text{ for } j\in\{1,...,i-1\} \text{ and} \\
  \dist({\bm x},{\bm v}_j) \leq \dist({\bm x},{\bm v}_{j+1}) \text{ for } j\in\{i,...,p-1\}
 \end{array}
 \right\}\, .
\end{align}

Recall that the regular $(p-1)$-simplex $\Sr$ can be decomposed into $p!$ isometric orthoschemes thanks to the barycentric decomposition. 
Each vertex ${\bm v}_i$ of $\Sr$ is a vertex of $(p-1)!$ such orthoschemes. 
Let $R_i$ be the union of the $(p-1)!$ orthoschemes having ${\bm v}_i$ as a vertex. 
Then ${\bm x}\in R_i$ if and only if party $i$ is the agent's favorite party. 
Hence, a point ${\bm x}\in R_i$ represents a consistent opinion (in the sense described above) if and only if it satisfies the distance conditions from \eqref{def:consistency_polyhedron} for that $i$. 
For instance, $C\cap R_1=O$, the orthoscheme with vertices ${\bm v}_1,...,{\bm v}_{1,...,p}$ described above.

Recall that the permutations in $\mathcal{S}_{p}$ are in 1-1 correspondence with the orthoschemes in the barycentric decomposition of $\Sr$, which in turns provides a 1-1 correspondence with all possible ordering of the parties: the permutation $(i_1,...,i_{p})$ corresponds to the orthoscheme $\conv({\bm v}_{i_1},{\bm v}_{i_1,i_2},...,{\bm v}_{i_1,...,i_{p}})$, which corresponds to the opinion with party $i_1$ being the favorite party, party $i_2$ the next favorite party (since ${\bm v}_{i_1,i_2}$ is the midpoint of the edge $[{\bm v}_{i_1},{\bm v}_{i_2}]$), and party $i_{p}$ the least favorite party. 

Conversely, any party ordering corresponds to a permutation in $\mathcal{S}_{p}$.
Therefore, an opinion is consistent (in the sense described above) if and only if its corresponding permutation $(i_1,...,i_{p})$ satisfies the following conditions:
\begin{align}
 i_k < i_l < i_1 &\Rightarrow k>l\, , &&\text{ and } & i_k > i_l > i_1 &\Rightarrow k > l\, .
\end{align}	
In other words, in the sequence of indices $i_1,...,i_{p}$, the elements of the subsequences $i_1-1,...,1$ and $i_1+1,...,p$ must appear precisely in that order, but elements from different subsequences can be permuted. 
There are $\binom{p-1}{i_1-1}$ such permutations starting with $i_1$.
Hence, for all $i\in\{1,...,p\}$, $C\cap R_{i}$ consists in the $\binom{p-1}{i-1}$ orthoschemes corresponding to consistent opinions. 
In particular, $C\cap R_1$ is the orthoscheme $O_{1,...,p}$ and $C\cap R_{p}$ is the orthoscheme $O_{p,...,1}$.
It follows that the set $C$ of points in $\Sr$ corresponding to consistent opinions is a non-convex polyhedron, obtained as the union of 
\begin{align}
 \sum_{i=1}^{p}\binom{p-1}{i-1} &= 2^{p-1}\, ,
\end{align}
isometric orthoschemes described above. 
The vertices ${\bm v}_1,...,{\bm v}_{p}$ form a subset of the vertices of $C$, and the vertex ${\bm v}_i$ is a vertex of $\binom{p-1}{i-1}$ such orthoschemes.

In order for the opinion generation process to be unbiased, the probability to generate an opinion representative ${\bm x}\in C\cap R_i$ should be the same for all $i\in\{1,...,p\}$. 
The discussion above provides a natural way to generate such opinions. 
Observe that for each permutation $(i_1,...,i_{p})$, the point ${\bm v}_{i_1,...,i_{p}}$ is the center ${\bm c}$ of $\Sr$. 
For each vertex ${\bm v}_i$, move it along the ray $[{\bm v}_i,{\bm c}]$ towards ${\bm c}$ to the point ${\bm s}_i$ such that
\begin{align}
 \dist({\bm s}_i,{\bm c}) &= \frac{1}{\binom{p-1}{i-1}}\cdot\dist({\bm v}_i,{\bm c})\, . 
\end{align}
Notice that this procedure preserves the existing interfaces between $C\cap R_i$ and $C\cap R_j$ for all $i\neq j\in\{1,...,p\}$, so that the non-convex polyhedron $\Cr$ obtained that way can be seen a non-uniform radial retractation of $C$ with respect of ${\bm c}$.
The set $\Cr$ is the colored area in the right panel of Fig.~\ref{fig:ortho-slide}.

Because of Proposition~\ref{prop:vol}, the volumes of all orthoschemes building $C\cap R_i$ are reduced by a factor $\binom{p-1}{i-1}$, so that each of the resulting regions $\Cr\cap R_i$ has the same volume, which is equal to the volume of the isometric orthoschemes $O_{1,...,p}$ and $O_{p,...,1}$ (one has indeed that $C\cap R_1=\Cr\cap R_1$ and $C\cap R_p=\Cr\cap R_p$).

We summarize the process for generating a random opinion, drawn uniformly in the admissible opinion space: 
\begin{enumerate}
\item Take the barycentric decomposition of the standard regular simplex~$S_\text{reg}$ into $p!$ isometric orthoschemes.
\item From these orthoschemes, only consider the $2^{p-1}$ ones corresponding to consistent opinions (their union is the polyhedron~$C$).
\item For all $i\in\{1,...,p\}$, retract the orthoschemes in $C$ containing the vertex ${\bm v}_i$ (their union is the region $R_i$) radially with respect to the center ${\bm c}$ of $\Sr$ from a factor $\tbinom{p-1}{i-1}$. The union of these retracted orthoschemes forms the polyhedron $\Cr$. As a result, $\vol_{p-1}(\Cr\cap R_i)=\vol_{p-1}(\Cr\cap R_j)$ for all $i\neq j\in\{1,...,p\}$.
\item In $\Cr$, randomly pick $N$ points, where $N$ is the desired amount of opinions to be generated. The probability that a point $x\in \Cr$ belongs to the region $\Cr\cap R_i$ is the same for all $i\in\{1,...,p\}$.
\end{enumerate}
An implementation of this procedure is available online~\cite{repo}.

\section{Spectrum of the dynamics matrix}\label{sec:side-res}

{\bf Lemma.}
{\it 
 The matrix $M\coloneqq D^{-1}L + I_n$ [as defined in the Methods Section, Eq.~(5)] has a real positive spectrum.
}

\begin{proof}
 First, let us identify the eigenvalues of $D^{-1}L$, which are the zeros of the characteristic polynomial 
 \begin{align}
 \begin{split}
  0 &= \det\left(\lambda I_n - D^{-1} L\right) \\
  &= \det\left[D^{-1/2}\left(\lambda I_n - D^{-1/2} L D^{-1/2}\right)D^{1/2}\right] \\
  &= \det\left(\lambda I_n - D^{-1/2} L D^{-1/2}\right)\, .
 \end{split}
 \end{align}
 We then see that the spectrum of $D^{-1} L$ coincide with the spectrum of $D^{-1/2} L D^{-1/2}$. 
 The matrix $D^{-1/2} LD^{-1/2}$ being real symmetric, its spectrum (and consequently the spectrum of $D^{-1}L$) is real. 
 Furthermore, by definition of $M$, $\lambda$ is an eigenvalue of $D^{-1} L$ if and only if $\lambda+1$ is an eigenvalue of $M$. 
 The spectrum of $M$ is then real. 
 
 Second, by Gershgorin's Circles Theorem~\cite{Horn1994}, the eigenvalues of $M$ have a positive real part, which concludes the proof. 
\end{proof}


\section{Targeting electoral units}\label{sec:min_vs_rand}
\begin{figure*}[h]
 \centering
 \includegraphics[width=0.4\textwidth]{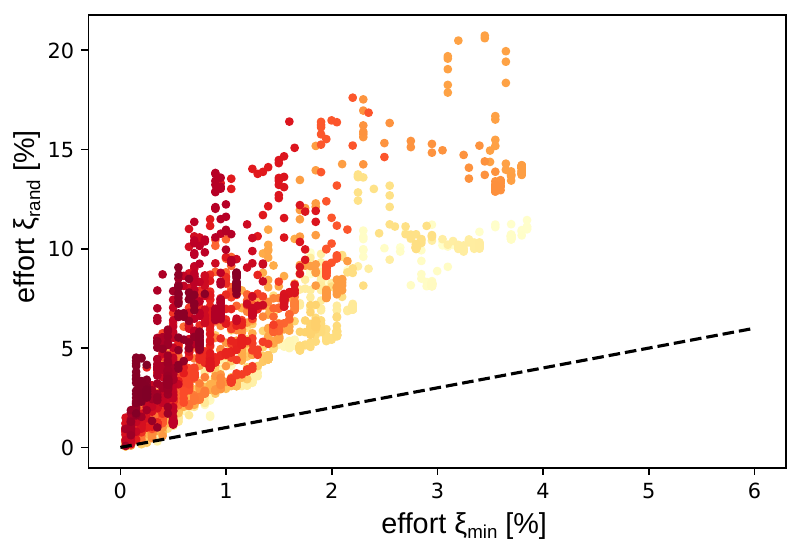}
 ~
 \includegraphics[width=0.4\columnwidth]{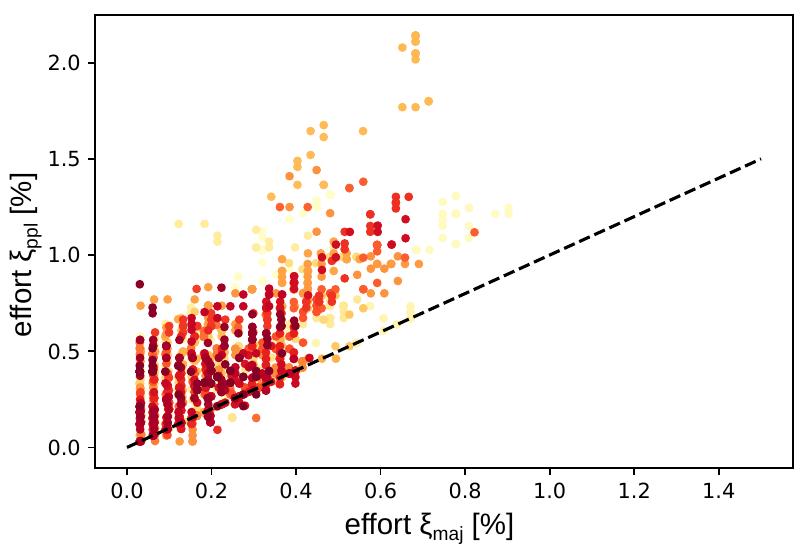}
 \caption{Left: The effort needed to change the outcome of election by targeting the agents close to the barycenter, $\xi_{\rm min}$, versus targeting random agents, $\xi_{\rm rand}$. 
 Each point denotes a natural opinion with different confidence bound $\epsilon$, polarization and bias.
 Right: The effort needed to change the outcome of election by targeting the electoral units with the minimal majority, $\xi_{\rm maj}$, and the electoral units with the minimal population, $\xi_{\rm ppl}$, with $7$ states. 
 We consider each state to have $400-500$ agents. 
 In both panels, different colors indicate different values of confidence bound $\epsilon$, polarization, and bias. }
 \label{fig:strategy}
\end{figure*}

When influencing agents in order to change the election outcome in an electoral unit, one would intuitively target the agents close to the center of the opinion distribution in priority as their vote is likely to be easier to change. 
This line of reasoning is confirmed in the left panel of Fig.~\ref{fig:strategy}, where we compare the effort needed to change the outcome of a vote by influencing the agents closer to the barycenter of the opinion space $\xi_{\min}$ to the effort when target agents at random $\xi_{\rm rand}$. 

When a country is composed of multiple electoral units, there are two natural ways to select which electoral unit to target in priority.
One can either target agents in the electoral unit with the lowest population, because changing the outcome in such a unit requires to influence less agents, or one can target the electoral unit with smallest relative majority, because such a unit is close to a change of outcome already. 
In the right panel of Fig.~\ref{fig:strategy}, we show that targeting the electoral units with lower relative majority is generally more efficient than targeting the electoral units with lower population.
Our strategy to overturn elections is therefore to target weakly opiniated agents in electoral units with lower population.

\section{Model Validation and the US House of Representatives Election}
For the validation of our model, we compare historical and numerical volatility of the House of Representatives elections in the US, between 2012 and 2020~\cite{us-election-results}. 

We measure the historical volatility $V_{\rm h}$ of the House of Representatives election in each district by counting the number of times the majority (Democrats or Republicans) has changed from $2012$ to $2020$.
A single representative is elected at the level of the districts for the House of Representatives elections. 
The more this number changes over time, the more volatile the electoral unit is. 
This volatility measure is illustrated in Fig.~1a.
Over the period $2012-2020$, $78$ districts changed majority at least once (Fig.~1a in the main text). 

\begin{figure*}[h]
 \centering
 \includegraphics[width=\textwidth]{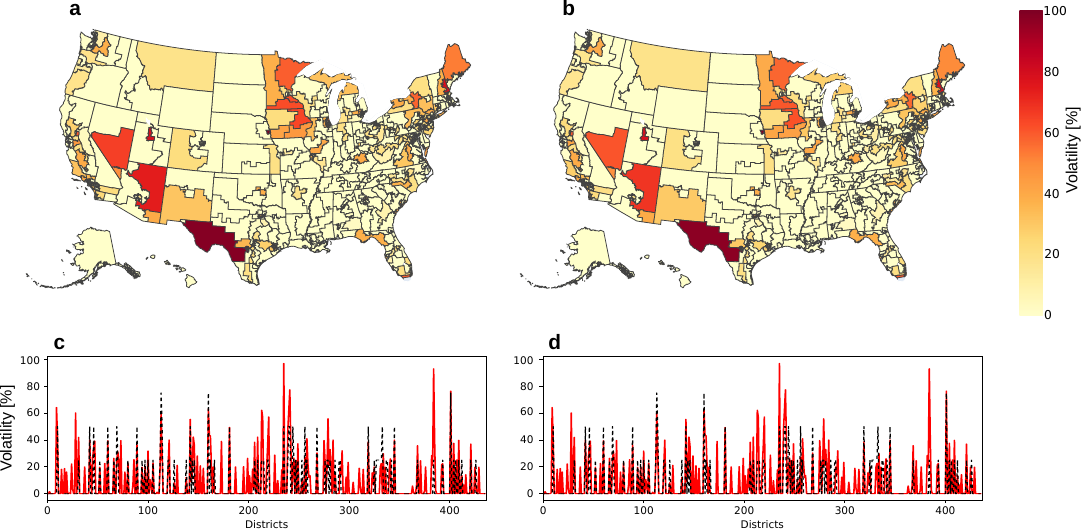}
 \caption{{\bf Supplementary figure:} 
 Color scale of the simulated volatility {\bf a} for a bigaussian distribution with bias and {\bf b} for a gaussian distribution with bias. 
 Panel {\bf c} (resp. {\bf d}) shows the line plot of the above color scale in panel {\bf a} (resp. {\bf b}) where districts are arranged along the horizontal axis. 
 The dashed black line shows the historic volatility and the plain red line shows the simulated volatility. 
 }
 \label{fig:validation-d2-d3}
\end{figure*}

To compute the numerical volatility $V_{\rm t}$ given by our model, we attribute an \emph{effort budget} $\Xi\in[0.0025,0.2]$ to influence a certain percentage of agents in the country, which is distributed evenly among the electoral units, i.e., each electoral unit, we can influence up to a fraction $\Xi$ of the agents. 
We then simulate the election in each district with $n_{\rm d}=501$ agents each and compute the number of times this influence can change the outcome in each of these electoral units over $100$ realisations of natural opinions for each type of distribution described in Sec.~C. 
After influence, each district may or may not change the majority.
Averaging overall realisations gives our estimate of volatility. 

We calculate the Pearson correlation coefficient between the numerically computed volatility and the historical volatility from the $2012$ to $2020$ US House of Representatives elections.

\section{Robustness of Bipartite Systems in Single Electoral Units} 

\begin{figure*}[h]
 \centering
 \includegraphics[width=0.9\textwidth]{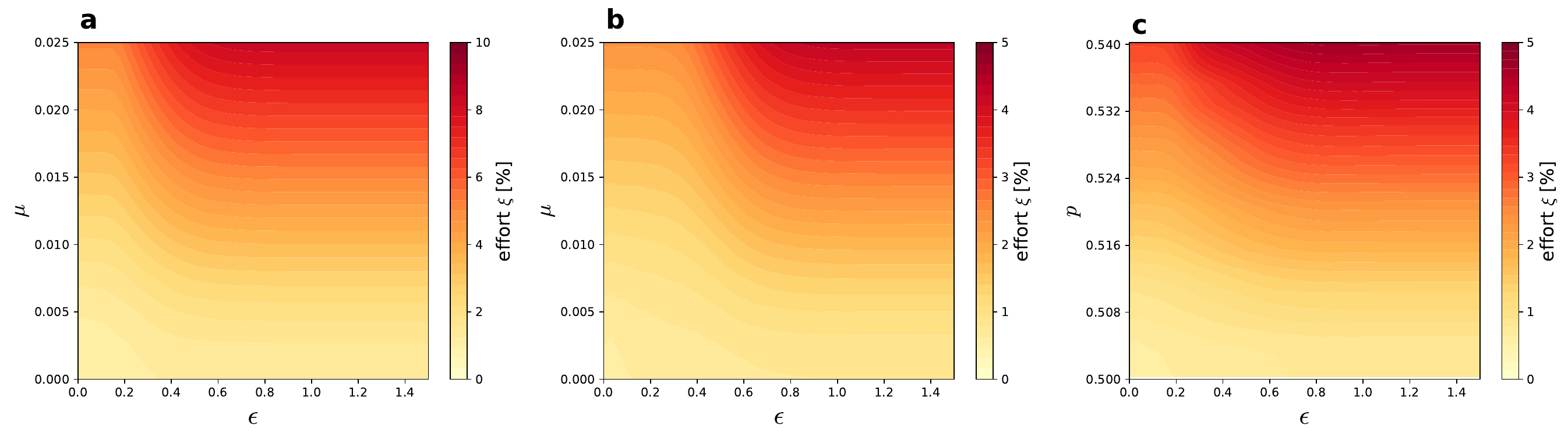}
 \caption{{\bf Supplementary figure:} 
 Average effort needed to change the outcome of an election with respect to different parameters of the opinion distributions of Eq.~(12) and the confidence bound $\epsilon$. 
  The average is taken over $500$ realisations of \emph{natural opinion} with $2001$ agents in an electoral unit for each simulation.
  {\bf a} Average effort as a function of the confidence bound $\epsilon$ and the distribution bias $\mu$, for a gaussian distribution ($\Delta=0.0$, $\sigma=0.2$, and $\rho = 0.5$). 
  {\bf b} Average effort as a function of the confidence bound $\epsilon$ and the distribution bias $\mu$, for a bigaussian distribution ($\Delta=0.5$, $\sigma = 0.2$, and $\rho = 0.5$). 
  {\bf c} Average effort as a function of the confidence bound $\epsilon$ and the balance parameter $\rho$, for a bigaussian distribution ($\Delta = 0.5$, $\sigma = 0.2$, and $\mu = 0.0$). 
  }
 \label{fig:variation}
\end{figure*}

In the main text we mentioned that biases parameters $\mu$ and $\rho$ of Eq.~(12) result in clear-cut electoral results, i.e. where there is a
substantial percentage difference between the winner and the first runner-up. It is then expected that such a result is robust against opinion
manipulation, because the opinion of a large numbers of voters need to be changed in that case. 
Fig.~\ref{fig:variation} confirms this expectation -- increasing the electoral bias. with either $\mu$ or $\rho$ increases the effort needed to change
the electoral outcome. 
As mentioned in the main text, one concludes, rather trivially, that population with electoral biases are robust against computational propaganda.


\section{Gerrymandering} 

Geographical redistricting may change the outcome of an election in favor of one or another party in a state with several districts. 
Focusing on bipartite elections, the goal is to achieve a slight majority in favor of one party in as many districts as possible, while concentrating 
the votes of the other party in as few districts as possible. This procedure is called gerrymyndering and we briefly consider its impact on electoral 
robustness in bipartite elections. 

\begin{figure*}[h]
 \centering
 \includegraphics[width=.6\textwidth]{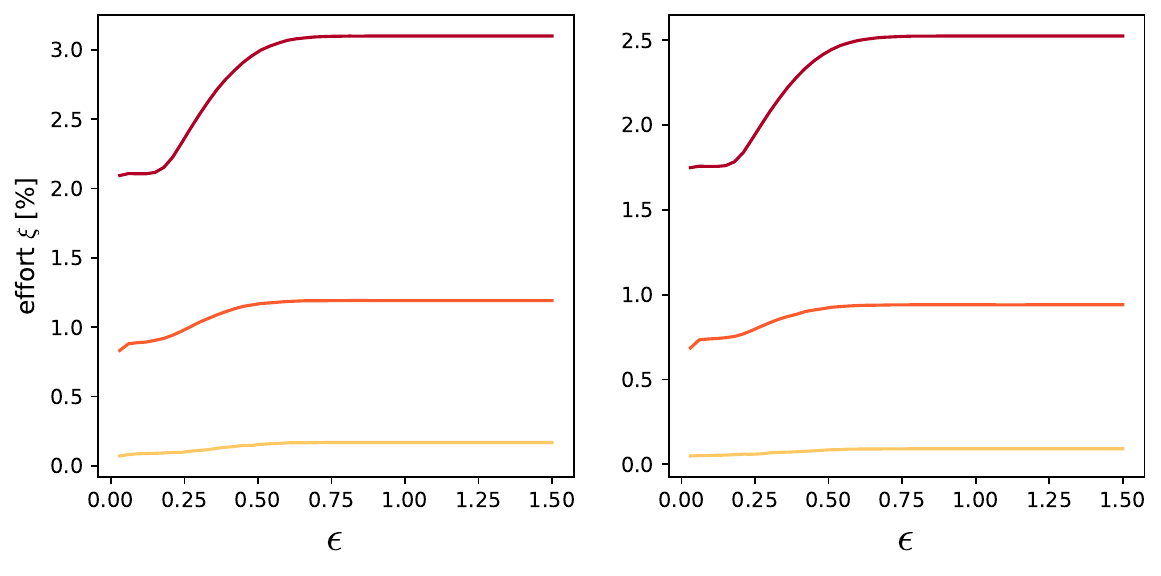}
 \caption{{\bf Supplementary figure:} 
 Average effort needed to change the outcome of a bipartite election in a state with five (left) and nine (right) electoral districts 
 and  for opinion distributions given in Eq.~(12) with $\sigma=0.2$, $\mu=0$, $\rho=0$ and $\Delta=0$.
 Different levels of gerrymandering are considered: no gerrymandering where natural opinions are distributed
 equally between the two parties in each district (yellow curve), where redistricting leads to 52 \% vs 48 \% 
 of natural opinions in all but one districts and 42 \% vs. 58 \% in the last one (orange), and to 55 \% vs. 45 \% in all but one districts
 and 30 \% vs 70 \% in the last one (red).
  }
 \label{fig:gerry}
\end{figure*}

In Fig.~\ref{fig:gerry} we show that gerrymandered bipartite elections are more robust than those where electorates have homogeneously 
distributed natural opinions, equally distributed between the two parties.  This result is easily understood. Gerrymandering introduces 
an electoral bias in each district, therefore electoral outcomes have larger voting margins in favor of the winning party in 
each district. Many more voter opinions need to be reversed to change such electoral outcomes.

\section{Robustness of the House of Representatives elections}
\begin{figure*}[h]
 \centering
 \includegraphics[width=\textwidth]{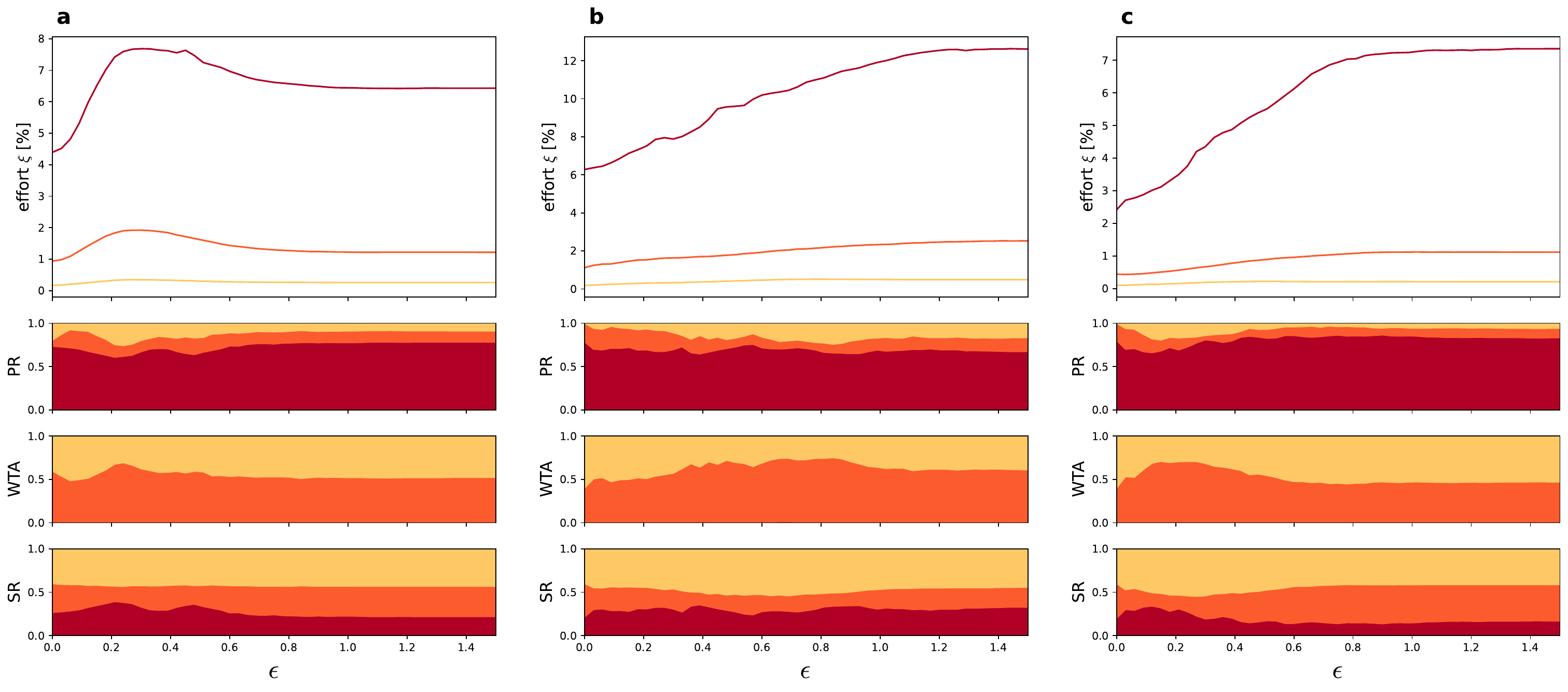}
 \caption{{\bf Supplementary figure:}
 Robustness evaluation on a simulation of the US House of Representatives elections, between 2012 and 2020 (100 realization for each election). 
 Elections are performed at the level of states, with results aggregated from the districts~\cite{us-election-results}. 
 Top row: average effort needed to change the election outcome for three electoral systems (PR: red, WTA: orange, and SR: yellow) and three distributions of natural opinions.
 Bottom rows: Proportion of times each electoral system was the most robust (red), second most robust (orange), and least robust (yellow). 
 }
 \label{fig:case_study_us}
\end{figure*}

In the main text (Fig. 3), we show how robustness evolves with respect to the opinion distance $\epsilon$ in randomly generated synthetic countries. 
Here we corroborate our findings on the US House of Representative elections between 2012 and 2020. 
We use the results of each of these elections as bias or shift in the opinion distribution. 
The simulations show little difference with the fully synthetic countries, except for the WTA system that appears to never be the most robust in this case (see Fig.~\ref{fig:case_study_us}). 

\section{Robustness for multipartite systems}
In the main text (Figs. 4 and 5), we show the robustness of the election outcome in a single electoral unit with 6 parties. 
Our finding is that the average effort needed to change an election is dominated by the effort needed to make the extremist parties win (when they are the first runner). 
A direct consequence if this observation is that, as soon as the extremist parties do not reach the second rank of the elections (which eventually happens when increasing the opinion distance), the robustness of the system drastically falls. 

\begin{figure*}[h]
 \centering
 \includegraphics[width=\textwidth]{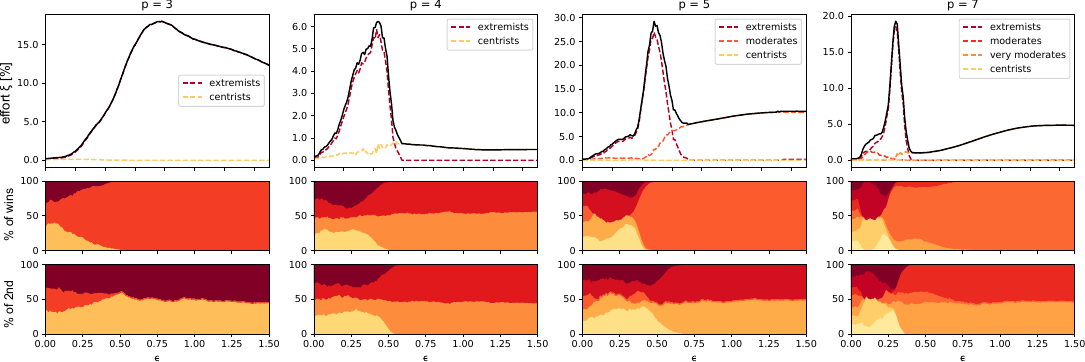}
 \caption{{\bf Supplementary figure:} 
 Same as Fig.~4 in the main text, for 3, 4, 5, and 7 parties (from left to right). 
 Top row: average effort to change the election outcome, as a function of the confidence bound $\epsilon$. 
 The effort is broken down into the contribution of different realizations aggregated according to the position of the first runner-up in the political spectrum. 
 Middle row: Proportion of times each party wins the elections. 
 Bottom row: Proportion of times each party is the first runner. 
 }
 \label{fig:Per_of_win_runner_up}
\end{figure*}

\begin{figure*}[h]
 \centering
 \includegraphics[width=\textwidth]{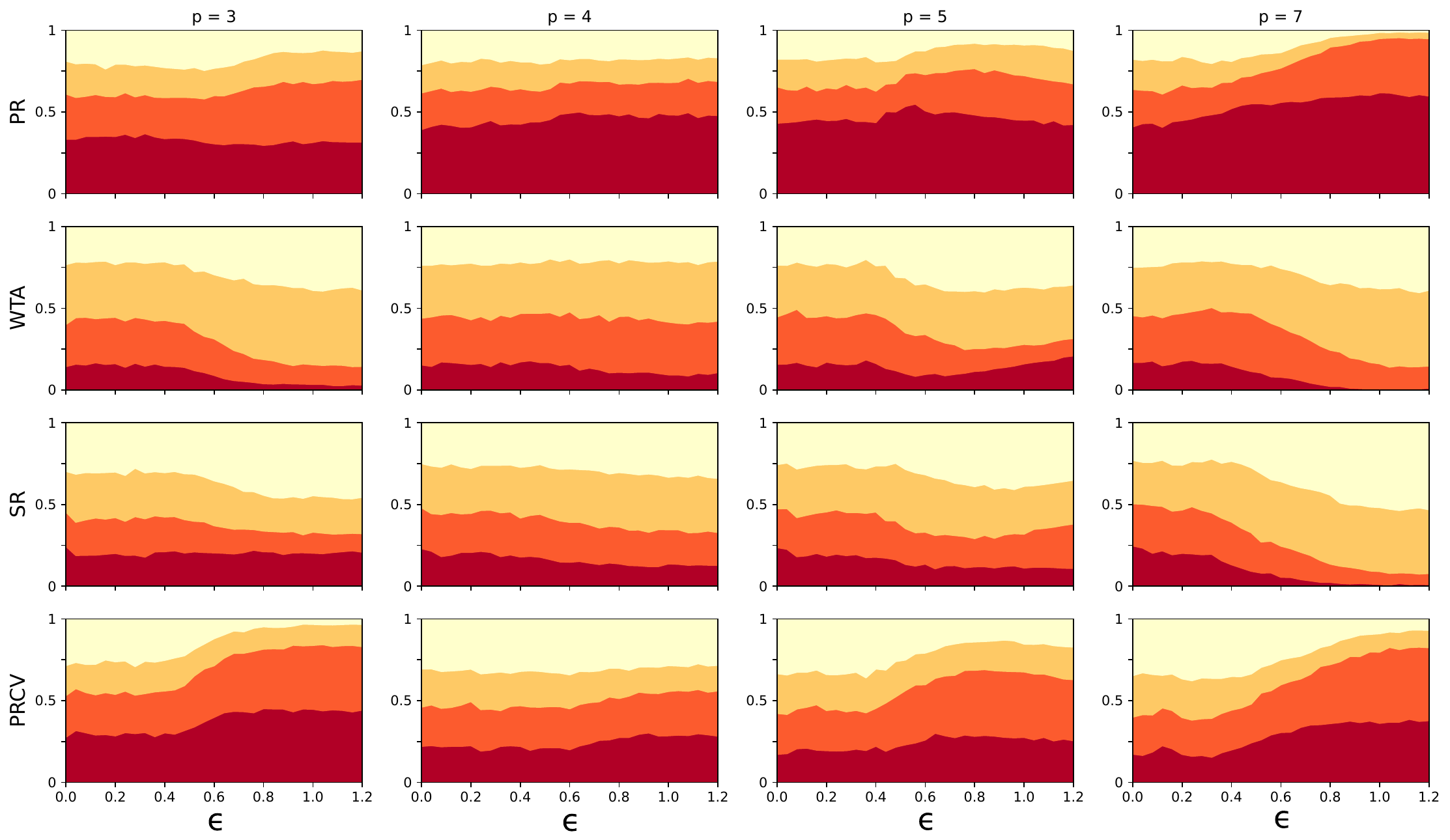}
 \caption{{\bf Supplementary figure:} 
 Same as Fig.~5 in the main text, for 3, 4, 5, and 7 parties (from left to right) and for the four electoral systems: PR, WTA, SR, and PRCV (from top to bottom). 
 Proportion of realizations where the electoral system is the most robust (dark red), second most robust (orange), and least robust (yellow) in our simuliations. 
 }
 \label{fig:fig5sup}
\end{figure*}

We show in Figs.~\ref{fig:Per_of_win_runner_up} and \ref{fig:fig5sup} these observations are valid a number of parties ranging from 3 to 7, without noticeable differences. 

\end{document}